\documentclass[twocolumn]{aa}
\usepackage[colorlinks,linkcolor={blue},citecolor={blue},urlcolor={red}]{hyperref}
\usepackage{threeparttable}
\usepackage{graphicx}
\usepackage{placeins}
\usepackage{float}
\usepackage{url}
\usepackage{natbib}
\usepackage{xspace}

\def\apjs{ApJS}

\def\beq{\begin{equation}}
\def\eeq{\end{equation}}
\def\bey{\begin{eqnarray}}
\def\eey{\end{eqnarray}}

\def\Mpc{\,{\rm Mpc}}
\def\Kpc{\,{\rm Kpc}}
\def\mpc{\, h^{-1}{\rm {Mpc}}}

\def\kms{\,{\rm {km\, s^{-1}}}}
\def\Msun{{\rm M_\odot}}

\def\gs{\mathrel{\raise1.16pt\hbox{$>$}\kern-7.0pt
\lower3.06pt\hbox{{$\scriptstyle \sim$}}}}
\def\ls{\mathrel{\raise1.16pt\hbox{$<$}\kern-7.0pt
\lower3.06pt\hbox{{$\scriptstyle \sim$}}}}
\def\gtsima{\, {\buildrel > \over \sim} \,}
\def\ltsima{\, {\buildrel < \over \sim} \,}
\def\prosima{\, {\buildrel \propto \over \sim} \,}
\def\gsim{\lower.5ex\hbox{\gtsima}}
\def\lsim{\lower.5ex\hbox{\ltsima}}
\def\simgt{\lower.5ex\hbox{\gtsima}}
\def\simlt{\lower.5ex\hbox{\ltsima}}
\def\simpr{\lower.5ex\hbox{\prosima}}

\begin{document}

\title{Massive Star-Forming Galaxies Have Converted Most of Their Halo Gas into Stars}
\author{Ziwen Zhang\inst{1,2}, Huiyuan Wang\inst{1,2}, Wentao Luo\inst{1,2},  Jun Zhang\inst{3}, H. J. Mo\inst{4}, YiPeng Jing\inst{3,1}, Xiaohu Yang\inst{3}, and  Hao Li\inst{1,2}}

\titlerunning{High Efficiency in Massive Star-forming Galaxies}
\authorrunning{Zhang et al.}

\institute{CAS Key Laboratory for Research in Galaxies and Cosmology, Department of Astronomy, University of Science and Technology of China, Hefei, Anhui 230026, China; Email: ziwen@mail.ustc.edu.cn, whywang@ustc.edu.cn
 \and School of Astronomy and Space Science, University of Science and Technology of China, Hefei 230026, China
 \and Department of Astronomy, and Tsung-Dao Lee Institute, Shanghai Jiao Tong University, Shanghai 200240, China
 \and Department of Astronomy, University of Massachusetts, Amherst MA 01003-9305, USA}

\abstract{
In the local Universe, the efficiency for converting baryonic gas into 
stars is very low. In dark matter halos where 
galaxies form and evolve, the average efficiency varies 
with galaxy stellar mass and has a maximum of about twenty percent for 
Milky-Way-like galaxies. The low efficiency at higher mass is believed to 
be produced by some quenching processes, 
such as the feedback from active galactic nuclei. 
We perform an analysis of weak lensing and satellite kinematics for SDSS central galaxies. 
Our results reveal that the efficiency is much higher, more than sixty 
percent, for a large population of massive star-forming galaxies around $10^{11}\Msun$.
This suggests that these galaxies acquired most of the gas in their halos 
and converted it into stars without being affected significantly by quenching processes. 
This population of galaxies is not reproduced in current galaxy 
formation models, indicating that our understanding of galaxy formation 
is incomplete.  The implications of our results on circumgalactic 
media, star formation quenching and disc galaxy rotation curves are discussed.
We also examine systematic uncertainties in halo-mass and stellar-mass 
measurements that might influence our results.
}

\maketitle

\section{Introduction}

In the standard $\Lambda$ cold dark matter cosmogony, galaxies 
are believed to form and evolve within dark matter halos. The
baryonic gas in halos cools radiatively and condenses and is then converted into stars\citep[][]{WhiteRees1978, FallEfstathiou1980}. The global efficiency for converting baryonic gas into 
stars is very low \citep{Bregman2007}, about 10\%. 
Within halos, the efficiency is usually defined as $M_*/M_{\rm h}/f_{\rm b}$, where $M_*$, $M_{\rm h}$ and $f_{\rm b}$ are stellar mass, halo mass and cosmic mean baryon fraction, respectively. So the efficiency is equivalent to the stellar mass-halo mass relation (SHMR).
It is found that the efficiency varies strongly with the halo mass and stellar mass of central galaxies, which are the dominant galaxies in the halos.
The efficiency reaches a maximum of about twenty percent for 
Milky-way-like galaxies and declines quickly towards the lower and higher masses \citep[e.g.][]{Yang2003,Yang2009, Leauthaud2012, Moster2013MNRAS.428.3121M, 
LuZhankui2014,
Hudson2015, Mandelbaum2016, Wechsler2018, Behroozi2019}. 

The efficiency is expected to be the result of many physical processes in and around galaxies, 
and its mass-dependence may reflect the relative importance of individual processes.
For example, the decline at the low-mass side may be produced by supernova feedback 
and stellar winds \citep[e.g.][]{Kauffmann1998MNRAS.294..705K, Cole2000MNRAS.319..168C}. 
The gravitational potential wells associated with these low-mass galaxies 
are expected to be shallow, so that these processes can effectively 
prevent star formation and the growth of galaxies. In contrast, 
at the high mass end, where effects of supernova feedback may not be
important, the suppression of the star formation efficiency
is usually believed to be caused by the energetic feedback from active galactic 
nuclei \citep[AGNs, e.g.][]{Silk1998, Croton-06, Fabian-12, Heckman2014}, 
although other processes might also be at work, such as morphological quenching 
and virial shock heating \citep[e.g.][]{Dekel2006, Martig-09}.

The low efficiency described above is the average for galaxies 
of a given stellar mass. It is clearly interesting to check whether the 
efficiency varies with galaxy properties other than the stellar mass. 
There is a growing amount of evidence that red/quiescent/early type galaxies reside 
in more massive halos than blue/star-forming/late type galaxies of the same 
stellar mass \citep[e.g.][]{More2011, Rodriguez-Puebla2015ApJ...799..130R, Mandelbaum2016, Behroozi2019, Lange2019,  Bilicki2021, Posti2021, Xu2022, Zhang2021}, indicating that the efficiency
for star-forming galaxies is higher than that for quiescent galaxies. 
Moreover, the efficiency also seems to be related to galaxy morphology \citep[e.g.][]{Mandelbaum2006, Xu2022}. 
For example, \cite{Posti2019} found the efficiency for disc galaxies increases monotonously 
with increasing stellar mass and deviates significantly from
that for the total galaxy population at the massive end.
More recently, \cite{Zhang2021} found that the host galaxies of optical AGNs
have stellar to halo mass ratio
(hereafter SHMR) similar to star-forming galaxies but different from quiescent galaxies. 
The mean halo mass of these AGNs is around $10^{12}\Msun$ \cite[see also][]{Mandelbaum09}, 
where the star formation efficiency peaks. This hints that AGNs 
tend to be triggered in galaxies with high star formation efficiency. 
Another interesting finding is that the peak efficiency declines, 
and the peak position shifts to lower mass as cosmic time 
progresses \citep[e.g.][]{Hudson2015}. 
Although other works using very different approaches gave different results \citep[e.g.][]{Moster2013MNRAS.428.3121M, LuZhanKui2015, Behroozi2019}.

One important step to evaluate the efficiency is to measure the halo mass for a galaxy sample
or for individual galaxies. In the literature, many methods have been 
developed to infer the halo masses from observational data, such as weak 
lensing, satellite kinematics, rotational velocity, galaxy clustering, galaxy abundance, X-ray emission 
and the Sunyaev-Zel'dovich effect (SZ effect). Galaxy-galaxy lensing and satellite kinematics are two powerful 
tools to measure halo mass, and have been investigated in great details \citep[e.g.][]{vandenBosch2004, Mandelbaum2006, Conroy07apj, More2011, Uitert2011, Leauthaud2012, Tinker2013ApJ, Wojtak2013, Velander2014MNRAS.437.2111V, Hudson2015, Viola2015MNRAS.452.3529V, Zu2015MNRAS, Mandelbaum2016, Shan2017, Luo2018ApJ, Lange2019a, Zhang2021}. 
In this paper, we combine the data of both galaxy-galaxy lensing and satellite kinematics 
to measure halo masses of galaxy samples with different stellar mass and 
star formation rate, and then to evaluate the efficiency of converting baryonic 
gas into stars for those samples. We also check our results by using galaxy clustering. 

The paper is organized as follows. Section \ref{sec_sp} presents the sample selection and
our method of using lensing and satellite kinematics to infer halo mass. 
In Section \ref{sec_res}, we show our main results for different galaxies,  
test uncertainties, and make comparisons with other results. 
We discuss the implications of our results in Section \ref{sec_dis}. 
Finally, we summarize our results in Section \ref{sec_sum}.
Throughout this paper, we assume the Planck Cosmology \citep{Planck2016}:
$\Omega_{\rm m}=0.307$, $\Omega_{\rm b}=0.048$, $\Omega_{\Lambda}=0.693$, 
$\rm h=0.678$. 
The cosmic mean baryon fraction is $f_{\rm b}=\Omega_{\rm b}/\Omega_{\rm m}=0.157$. 

\section{Samples and Methods of Analysis}

\subsection{Sample properties}\label{sec_sp}

Our galaxy sample is selected from the New York University Value Added 
Galaxy Catalog (NYU-VAGC) \citep{Blanton-05a} of 
the Sloan Digital Sky Survey (SDSS) DR7 \citep{Abazajian-09}. 
Galaxies with $r$-band Petrosian magnitudes $r\le$ 17.72, with redshift 
completeness $>$ 0.7, and with redshift range of 
$0.01 \le z \le 0.2$, are selected.
We only focus on central galaxies that are the most massive galaxies within dark matter halos, 
and the halo-based group catalog of \citet{Yang2007} is adopted to identify centrals 
\footnote{The galaxy group catalog is publicly available at https://gax.sjtu.edu.cn/data/Group.html}. 
NYU-VAGC provides measurements of the sizes of a galaxy, $R_{50}$ and $R_{90}$,
which are the radii enclosing 50 and 90 percent of the Petrosian $r$-band flux
of the galaxy, respectively \footnote{The galaxy size data can be downloaded at http://sdss.physics.nyu.edu/vagc/}. The concentration of a galaxy, defined 
as $C=R_{90}/R_{50}$, is usually used to
indicates the morphology of the galaxy \citep{Shimasaku2001, Strateva2001}. 
It also provides colors, e.g. $(g-r)^{0.1}$, and the S$\acute{e}$rsic radial 
profile fits for galaxies.

We cross-match our sample with the MPA-JHU DR7 catalog
\footnote{https://wwwmpa.mpa-garching.mpg.de/SDSS/DR7/} to obtain the measurements of
stellar mass ($M_*$) \citep{Kauffmann2003} and star formation rate (SFR) \citep{Brinchmann2004}.
The stellar mass is obtained by fitting the SDSS $ugriz$ photometry
to models of galaxy spectral energy distribution (SED), 
and the results are excellent agreement with those obtained 
by \citet{Moustakas2013} using photometry in 12 UV, optical and infrared 
bands. The SFR is derived using both the spectroscopic and photometric 
data of the SDSS. We define star-forming galaxies as the ones on the 
star-formation main sequence given by $\log({\rm SFR})=0.73\log M_*-7.3$
\citep{Bluck2016}. The dispersion of the main sequence is about 0.3 dex \citep[e.g.][]{Speagle2014,Kurczynski2016}. We, therefore, identify 
galaxies above $\log({\rm SFR})=0.73\log M_*-7.6$ as star-forming galaxies. 

As shown below, we will use the weak lensing shear catalog measured from
the Dark Energy Camera Legacy Survey \citep[hereafter DECaLS][]{Dey2019AJ} 
to measure the halo masses of our galaxies. We thus only focus on galaxies within 
the DECaLS region. This selection excludes about thirty-two percent of galaxies.
In this way we select two samples of central galaxies, one for star-forming galaxies,
which consists of 129,278 galaxies with $M_*\ge 10^{8.8}\Msun$, and the other 
for the total population, which consists of 304,162 galaxies in the same 
mass range and is used for comparison. We divide each of the two samples 
into six subsamples, equally spaced in the logarithm of stellar mass with 
a bin size of 0.5 dex. The second most massive subsample, which is of particular 
importance for our analysis,  has a mean stellar mass of $M_*\sim 10^{11}\Msun$ 
and contains a total of 106,125 galaxies, including 22,099 star-forming galaxies. 
See Table \ref{table1} for the stellar mass range, the mean stellar mass, and 
the galaxy number for each of the subsamples.

\begin{table*}
\centering
\caption{The properties of star-forming and total galaxy samples in the DECaLS region.}\label{table1}
\begin{threeparttable}
\resizebox{\textwidth}{!}{
\begin{tabular}{c c c c c c c c}
\hline\
$\log \rm M_*$ range\tnote{(a)} & $\log \rm \bar M_*/{\rm M_\odot}$\tnote{(b)} & N$_{\rm c,gal}$\tnote{(c)} & N$_{\rm s,gal}$\tnote{(d)} & $\log \rm M_{\rm h}/{\rm M_\odot}$\tnote{(e)} & Efficiency\tnote{(f)} & LcSK Mass\tnote{(g)} & LcSK Eff.\tnote{(h)} \\
\hline
 & & & & Star-forming\\\relax
[8.84, 9.34] & 9.14 & 7,940 & 213 & 10.99$^{+0.43}_{-0.51}$ & 0.089$^{+0.197}_{-0.056}$ & 11.27$^{+0.37}_{-0.32}$ & 0.047$^{+0.052}_{-0.027}$\\\relax
[9.34, 9.84] & 9.64 & 16,903 & 756 & 11.44$^{+0.22}_{-0.35}$ & 0.102$^{+0.127}_{-0.04}$ & 11.13$^{+0.11}_{-0.11}$ & 0.207$^{+0.061}_{-0.046}$\\\relax
[9.84, 10.34] & 10.13 & 33,826 & 1,452 & 11.57$^{+0.11}_{-0.14}$ & 0.233$^{+0.087}_{-0.053}$ & 11.56$^{+0.06}_{-0.07}$ & 0.241$^{+0.042}_{-0.031}$\\\relax
[10.34, 10.84] & 10.61 & 47,759 & 2,829 & 11.92$^{+0.06}_{-0.07}$ & 0.308$^{+0.05}_{-0.04}$ & 11.89$^{+0.04}_{-0.04}$ & 0.335$^{+0.034}_{-0.031}$\\\relax
[10.84, 11.34] & 11.03 & 22,099 & 1,272 & 12.03$^{+0.08}_{-0.09}$ & 0.629$^{+0.142}_{-0.109}$ & 12.01$^{+0.05}_{-0.04}$ & 0.66$^{+0.069}_{-0.067}$\\\relax
[11.34, 11.82] & 11.43 & 749 & 85 & 12.45$^{+0.22}_{-0.35}$ & 0.61$^{+0.755}_{-0.238}$ & 12.47$^{+0.35}_{-0.25}$ & 0.58$^{+0.454}_{-0.32}$\\
 \hline
 & & & & Total \\\relax
[8.84, 9.34] & 9.14 & 8,984 & 357 & 11.15$^{+0.42}_{-0.57}$ & 0.063$^{+0.169}_{-0.039}$ & 11.02$^{+0.15}_{-0.16}$ & 0.084$^{+0.036}_{-0.024}$\\\relax
[9.34, 9.84] & 9.64 & 19,777 & 1,417 & 11.63$^{+0.17}_{-0.23}$ & 0.066$^{+0.047}_{-0.021}$ & 11.49$^{+0.09}_{-0.08}$ & 0.09$^{+0.018}_{-0.017}$\\\relax
[9.84, 10.34] & 10.15 & 48,753 & 3,975 & 11.8$^{+0.09}_{-0.1}$ & 0.141$^{+0.036}_{-0.026}$ & 11.82$^{+0.04}_{-0.05}$ & 0.135$^{+0.016}_{-0.012}$\\\relax
[10.34, 10.84] & 10.64 & 105,577 & 15,408 & 12.2$^{+0.04}_{-0.04}$ & 0.175$^{+0.018}_{-0.016}$ & 12.19$^{+0.02}_{-0.02}$ & 0.178$^{+0.008}_{-0.009}$\\\relax
[10.84, 11.34] & 11.09 & 106,125 & 27,873 & 12.59$^{+0.03}_{-0.03}$ & 0.198$^{+0.013}_{-0.012}$ & 12.62$^{+0.02}_{-0.02}$ & 0.186$^{+0.008}_{-0.007}$\\\relax
[11.34, 11.84] & 11.47 & 14,946 & 16,584 & 13.37$^{+0.02}_{-0.02}$ & 0.08$^{+0.004}_{-0.004}$ & 13.32$^{+0.03}_{-0.03}$ & 0.09$^{+0.007}_{-0.007}$\\
\hline
\end{tabular}}

\begin{tablenotes}
\footnotesize
\item[(a)] The stellar mass range.
\item[(b)] The mean stellar mass.
\item[(c)] The number of central galaxies.
\item[(d)] The number of satellite candidates.
\item[(e)] The halo mass obtained from galaxy-galaxy lensing.
\item[(f)] The efficiency obtained from galaxy-galaxy lensing.
\item[(g)] The halo mass ($\log \rm M_{\rm h}/{\rm M_\odot}$) obtained from lensing calibrated satellite kinematics method.
\item[(h)] The efficiency obtained from lensing calibrated satellite kinematics method.
\end{tablenotes}
\end{threeparttable}

\end{table*}

\subsection{Weak lensing measurements}
\label{sec_wlensing}

\begin{figure*}
\includegraphics[scale=0.8]{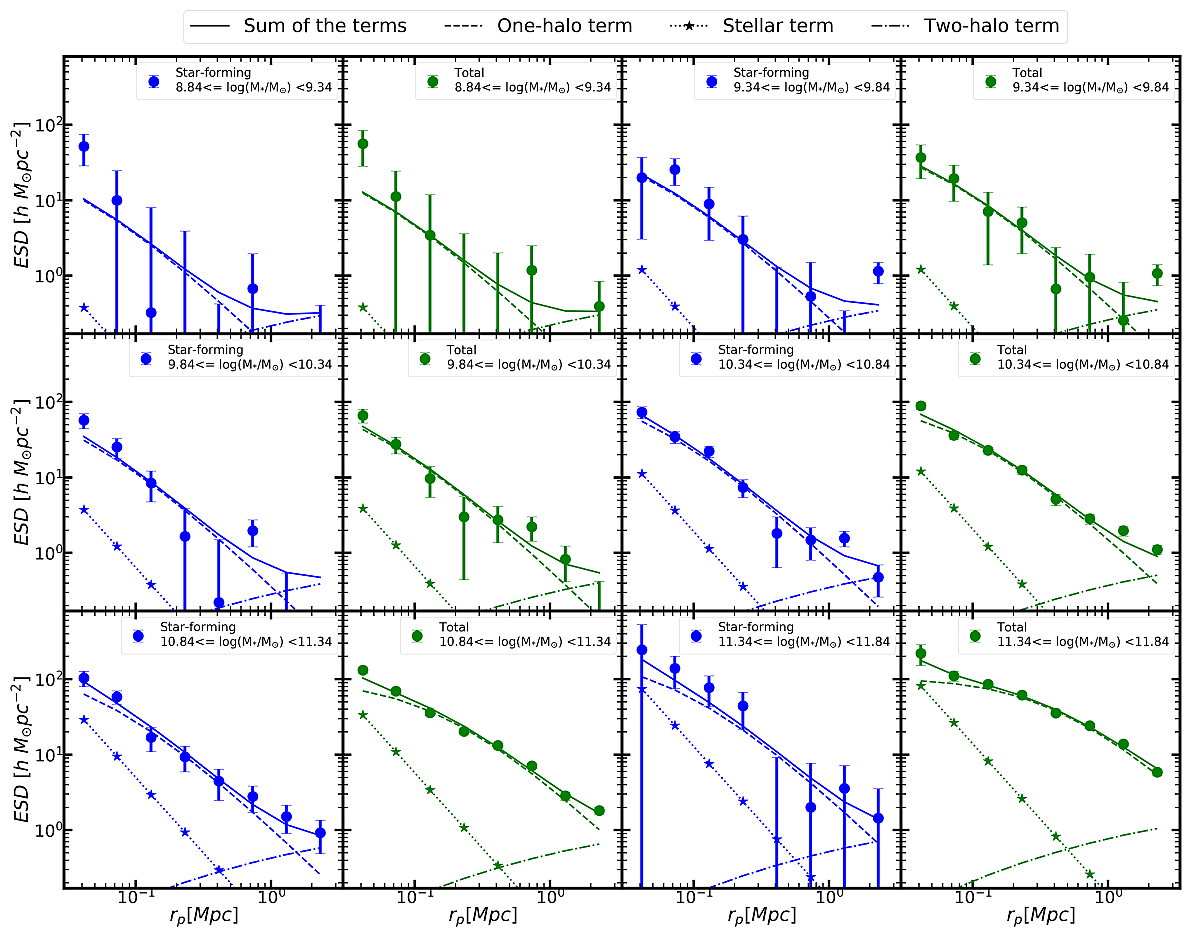}
\centering
\caption{Excess surface density from DECaLS shear catalog and the corresponding 
best fitting results. The symbols in twelve panels show the excess surface density (ESD) profiles obtained by stacking the lensing signals for galaxies 
in each stellar mass bin, as indicated in each panel. Results are shown separately 
for star-forming galaxies (1st and 3rd columns) and the total population 
(2nd and 4th columns).
The error bars correspond to the standard deviation of 150 bootstrap samples. 
We fit the ESD by using three components, the stellar mass term 
(dotted lines with stars), the one-halo term (dashed lines), and 
the two-halo term (dash-dotted lines). 
The sum of those components are shown by solid lines.}
\label{fig_lens}
\end{figure*}

\begin{figure*}
\includegraphics[scale=0.58]{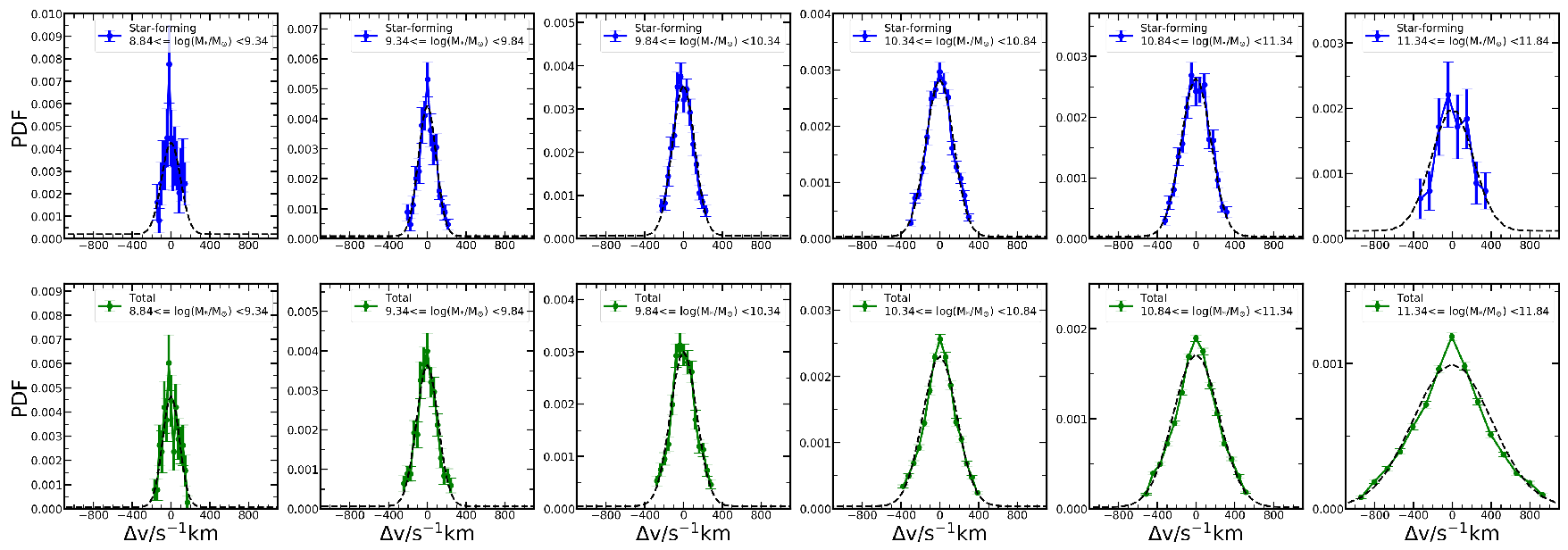}
\centering
\caption{The distribution of line-of-sight velocity difference ($\Delta v$) between central galaxies and their satellite candidates. The solid lines with error bars show the probability distribution functions (PDF) of $\Delta v$. The dashed lines show the Gaussian plus a constant fits to the data points.
The error bars correspond to the standard deviation of 100 bootstrap samples.
The upper-row shows the results for star-forming centrals, while the lower-row is for the total central populations. The results of star-forming/total samples in different stellar mass bins are shown in different columns, respectively. Note that the scales of the vertical 
axis are different for different panels.}\label{fig_deltav}
\end{figure*}

The shear catalog \footnote{The DECaLS shear catalog
is publicly available at https://gax.sjtu.edu.cn/data/DESI.html} used here 
to measure galaxy-galaxy lensing signals is based on the Dark Energy Camera 
Legacy Survey (DECaLS) DR8 imaging data \citep{Dey2019AJ,Zou_2019}.
The shape of each galaxy is measured using the FOURIER\_QUAD pipeline, which has been 
shown to yield accurate shear measurements even for extremely faint galaxy images 
(signal-to-noise ratio $< 10$) when applied to both simulations \citep{Zhang2015JCAP} 
and real data (\cite{Zhang_2019} for the CFHTLenS data, and \cite{Wanghr2021} for the DECaLS data). The whole shear catalog covers about 9,000 square
degrees in g/r/z bands, containing shear estimators from 190/246/300 
million galaxy images, respectively. Note that the images of the same galaxy in different exposures are counted as different images in the FOURIER\_QUAD method. 

Photometric redshifts of galaxies in the shear catalog are calculated 
using the random forest regression method \citep{Breiman2001}, a machine learning 
algorithm based on decision trees. Eight parameters were used in training the algorithm, 
including the $r$-band magnitude, $(g-r)$, $(r-z)$, $(z-W1)$, and $(W1-W2)$ colors, 
half-light radius, axial ratio, and shape probability. The photo-$z$ error is estimated  
for each individual shear galaxy by perturbing the photometry of the galaxy. 
Specifically, the uncertainty is assumed to follow a Gaussian distribution with the 
standard deviation equal to the photometric error; a random 
value generated from the distribution is added to the observed flux 
in each band to obtain a `perturbed' flux; the perturbation 
is repeated multiple times and the standard deviation of the photo-$z$ 
estimates from the perturbations is used as the error of the photo-$z$ 
\citep[see][for more details]{zhourongpu2021}.

Here we only use the $r$ and $z$ band data, because we find that the g-band images have some quality issues. The details of the image processing pipeline of the DECaLS data are given in another work (Zhang et al., in preparation). The overlapping region of DECaLS with SDSS DR7 is about 4744 square 
degrees. The estimator, Excess Surface Density (ESD), 
\begin{equation}
\rm \Delta \Sigma (R)=\gamma_t(R)\Sigma_{crit},
\end{equation}
is measured using the PDF-Symmetrization method \citep{Zhang2017}, which 
minimizes the statistical uncertainty. Note that due to the scatter as well as the uncertainty of the background galaxy redshifts, the PDF-Symmetrization method should be modified slightly. The details are given in a companion paper (Li et al., in preparation), which also includes a general discussion about different source of systematic errors in the measurement of the excess surface density within the framework of the PDF-Symmetrization method. In Figure \ref{fig_lens}, we show the 
ESD for both star-forming and total galaxy populations. 
The error bars are estimated by using 150 bootstrap samples
\citep{Barrow-Bhavsar-Sonoda-84}.

Following previous studies \citep[e.g.][]{Mandelbaum2008, Leauthaud2010, Luo2018ApJ}, 
we model the ESD using three components: 
\begin{equation}
 \rm \Delta \Sigma=\Delta\Sigma_{stellar}+\Delta\Sigma_{NFW}+\Delta\Sigma_{2h}.\\
\end{equation}
The first term is the contribution of the stellar mass of galaxies.
We adopt the stellar mass directly from the observational data 
and modeled it as a point mass. The second term is the contribution of
the dark matter halo, assumed to have a Navarro-Frenk-White\citep[NFW,][]{Navarro1997} 
density profile, described by two free parameters: the mass $m_{\rm h}$ and the concentration $c$.
Specifically, $m_{\rm h}$ is the mass contained in a spherical region of
radius $r_{\rm 200m}$, within which the mean mass density is equal to 
200 times the mean matter density of the Universe. 
The distributions of the halo mass and concentration 
for a fixed galaxy stellar mass are usually quite broad. 
In our modeling, we use a single NFW profile with two free parameters, 
$m_{\rm h}$ and concentration, to fit the data point, ignoring 
the dispersion in them. Since our analysis focuses only on the 
average information of halos, the bias produced by ignoring 
the dispersion is expected to be small, as shown 
in \cite{Mandelbaum2016} and to be discussed further in Section \ref{sec_tests}.
Following \cite{Mandelbaum2016}, we also ignore the effect of off-centering,
which is negligible for the $m_{\rm h}$ estimation of central galaxies according to our tests.
The third term, referred to as the two-halo term, is calculated by projecting 
the halo-matter cross correlation function, $\xi_{\rm hm}$, 
along the line-of-sight. Here $\xi_{\rm hm}=\rm b(m_{\rm h})\xi_{\rm mm}$, 
with $\rm \xi_{\rm mm}$ being the linear matter-matter correlation 
function and $\rm b(m_{\rm h})$ the linear halo bias \citep{Tinker2010}, 
both generated using \textit{COLOSSUS} \citep{Diemer2018ApJS}. 
We sample the posterior distribution
of the two parameters, $m_{\rm h}$ and $c$, using the Monte Carlo Markov Chain (MCMC) 
provided by a public open software, \textit{emcee} \citep{Foreman-Mackey2013PASP}. 
Both \textit{COLOSSUS} and \textit{emcee} are under the MIT License. 
The best-fitting profiles are presented in Figure \ref{fig_lens}. 
The quoted mass, $m_{\rm h}$, is the median value of the posterior and 
its error bar indicates the 16 and 84 percentage range of the posterior.

We note that $m_{\rm h}$ actually accounts for the contribution from cold dark matter, diffuse gas and satellites around centrals, 
but does not include the contribution from the central galaxies
that is modeled by $\Delta\Sigma_{\rm stellar}$.
The total mass of the halo, which is used to calculate the conversion 
efficiency in the following and should include all the components within 
the virial radius of the halo, is thus $M_{\rm h}=m_{\rm h}+M_*$. 
In general, $M_*\ll m_{\rm h}$ and so $M_{\rm h}$ is very close to $m_{\rm h}$. 
However, for halos with very high efficiency, central galaxies can 
also have a significant contribution to the total halo mass. In the rest of the paper, 
halo mass refers to the total mass of the halo, $M_{\rm h}$. 

\subsection{Weak lensing calibrated satellite kinematics method}\label{sec_skwl}

The kinematics of satellite galaxies provide an important 
probe of the gravitational potential wells of the dark matter halos
\citep[e.g.][]{McKay2002,vandenBosch2004,More2011, Wojtak2013, Lange2019a, Lange2019, Abdullah2020, Seo2020}. 
For a central galaxy of mass $M_*$ in a given subsample, we identify its satellite candidates from a reference galaxy sample. 
For our analysis, we define 
the reference sample as a magnitude-limited sample following the selection 
criteria: $r$-band Petrosian apparent magnitude of $r<17.6$, $r$-band Petrosian 
absolute magnitude in the range of $(-24, -16)$, and redshift in range of
$0.01 < z < 0.2$ \citep{WangL2019, Zhang2021}. 
The candidates are defined as the ones that satisfy the following criteria: 
$|\Delta v|\leq 3 v_{\rm vir}$, $r_{\rm p}\leq r_{\rm vir}$ and 
$M_{\rm s}< M_{\rm *}$. Here $r_{\rm p}$ and $\Delta v$ are the projected distance and 
the line-of-sight velocity difference between the central 
in question and it's satellite, respectively, $r_{\rm vir}$ and $v_{\rm vir}$ are, 
respectively, the virial radius and virial velocity calculated using 
the mean halo mass of the subsample derived from weak lensing, 
and $M_{\rm s}$ is the stellar mass of satellite. The numbers of satellite candidates for different central galaxy samples are listed in Table \ref{table1}.

Figure \ref{fig_deltav} shows the probability distribution functions (PDFs)
of $\Delta v$ for the selected satellites associated with different
(total/star-forming) central galaxy subsamples.
We use a Gaussian plus a constant, 
\begin{equation}
\frac{A}{\sqrt{2\pi}\sigma_v}{{\rm e}^{-\Delta v^2/2\sigma^2_v}}+d,\\    
\end{equation}
to fit the PDFs. Here, the Gaussian component represents the true satellites, 
and the constant component is used to account for the interlopers 
that are not physically associated with the centrals \citep[see also][]{McKay2002, Brainerd2003, vandenBosch2004, Conroy07apj}.
A MCMC technique is used to perform the fitting.
As can be seen from the figure, the $\Delta v$ distributions are well fitted by the two-component 
model and the contribution of the interlopers is negligible. 
Given that the uncertainty of a SDSS galaxy redshift is about 35$\kms$, 
the error of $\Delta v$ is $\sigma_e=\sqrt{2}\times35=49.5\kms$. 
So the velocity dispersion, $\sigma_{\rm s}$, of satellites can be 
estimated from the best-fitting Gaussian after correcting for 
the redshift uncertainties, $\sigma_{\rm s}=\sqrt{\sigma_v^2-\sigma_e^2}$. 
The estimate of $\sigma_{\rm s}$ may be affected by the uncertainty of the halo 
masses that are used to determine $r_{\rm vir}$ and $v_{\rm vir}$. 
As a check, for each sample we compare the $\sigma_{\rm s}$ calculated using three 
different halo masses, corresponding to the 16, 50 and 84 percentiles of the 
posterior distribution obtained from the MCMC fitting to the stacked lensing profiles
of the sample in question. We find that the three values of $\sigma_{\rm s}$ so obtained 
are in general consistent with each other (see Appendix \ref{Mh_uncertainty} for the 
detail). We thus conclude that the uncertainty in the halo mass 
has little impact on the estimate of $\sigma_{\rm s}$.

\begin{figure}
\includegraphics[scale=0.4]{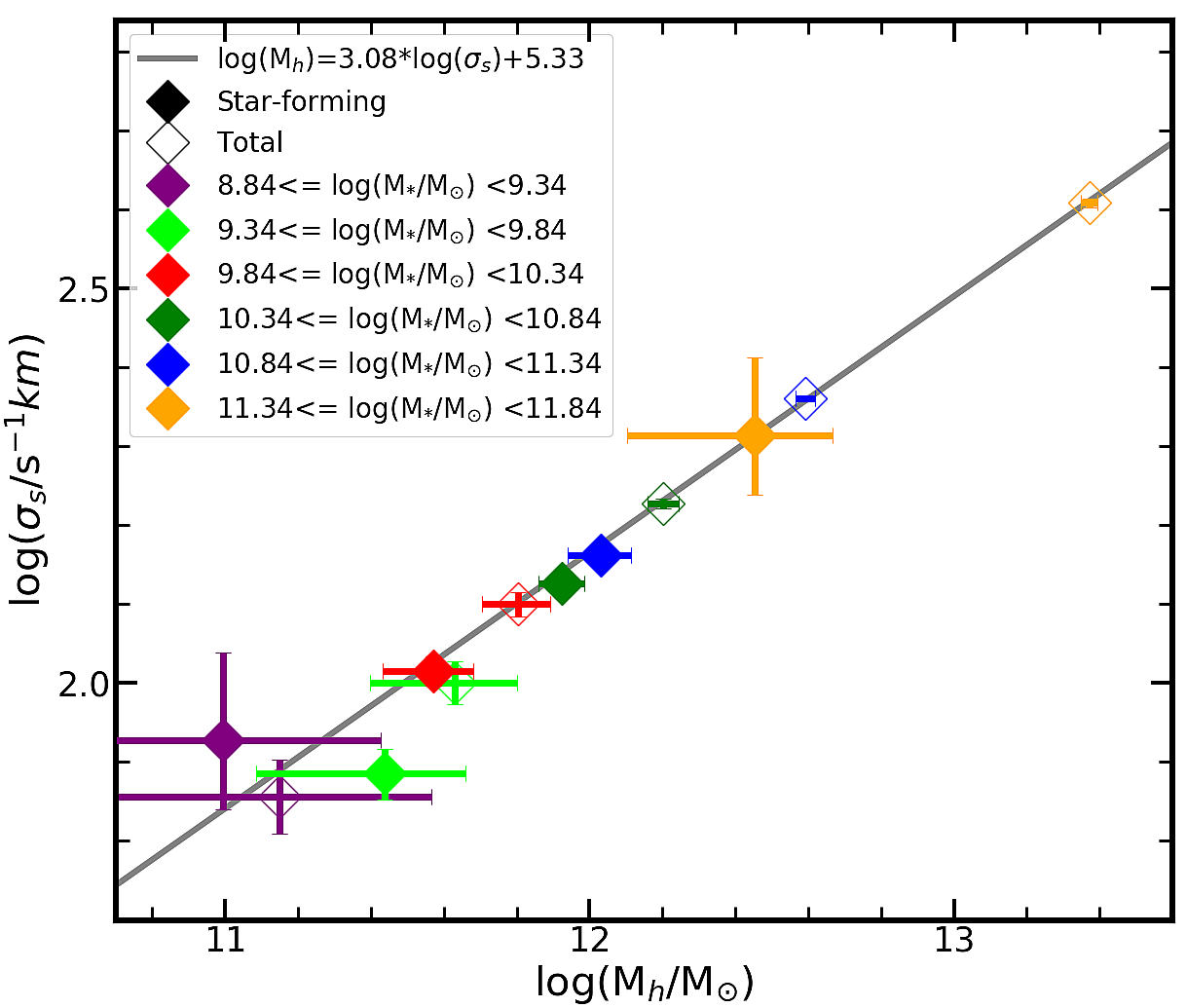}
\centering
\caption{Halo mass-satellite velocity dispersion relation. The open (solid) diamonds show the velocity dispersion ($\sigma_{\rm s}$) of satellites versus the host halo masses for total (star-forming) central galaxies. The halo masses are measured from DECaLS lensing data.
Different colors indicate different stellar mass bins of central galaxies.
The error bars for halo mass indicate the 16 and 84 percentiles of the posterior distribution obtained from the MCMC fitting to the stacked lensing profiles.
The error bars for $\sigma_{\rm s}$ represent the 16 and 84 percentiles of the posterior distribution obtained from the fitting to the distribution of the satellite-central velocity difference. 
The solid line represents the best-fitting to the data points for total galaxy sample. 
}\label{fig_sigmamh}
\end{figure}

In Figure \ref{fig_sigmamh}, we show $\sigma_{\rm s}$ versus $M_{\rm h}$, 
the lensing mass measurement, for both total and 
star-forming samples. One can see a strong correlation between the
two parameters. We fit the data points for the total sample with a power-law
model, and constrained relation is  
\begin{equation}\label{eq_mhsigma}
\log(\rm M_{\rm h}/\Msun) =(3.08\pm0.05)\log(\sigma_{\rm s}/\kms)+(5.33\pm0.11).\\     
\end{equation}
Remarkably, the slope of the relation is in excellent 
agreement with what is expected from the virial scaling relation 
($M\propto\sigma^3$) seen in numerical simulations\citep{Evrard2008}. 
A similar slope has been obtained in previous studies measuring halo mass 
using abundance matching, satellite kinematics, caustic technique, SZ effect, and 
virial theorem \citep[e.g.][]{Yang2007, More2011, Rines2013, Rines2016, Abdullah2020}, 
although some other studies found different results
\citep[e.g.][]{Viola2015MNRAS.452.3529V}. Our tests show that both weak lensing and
satellite kinematics can provide robust and consistent measurements of the host halo masses of galaxies.
The star-forming galaxies also closely follow the trend defined by the total sample, 
demonstrating that the lensing masses for star-forming galaxies are also robust. 

Thus, we can derive the host halo masses for the star-forming and total subsamples from their measured $\sigma_{\rm s}$ (Figure \ref{fig_deltav}) using the $\rm \sigma_{\rm s}$-$M_{\rm h}$ relation. But, $\sigma_{\rm s}$ is also used in fitting the relation, the uncertainties in $\sigma_{\rm s}$ and in the relation are expected to be correlated. In order to avoid this correlation, we design a new way to calculate the $\rm \sigma_{\rm s}$-$M_{\rm h}$ relation. Specifically, for each stellar mass bin, we only use data points ($\sigma_{\rm s}$ versus $M_{\rm h}$) in the rest of the stellar mass bins to fit the $\rm \sigma_{\rm s}$-$M_{\rm h}$ relation, so that the derived relation is independent of the galaxy sample in question. In Appendix \ref{new_Mh_sigma_relation}, we show the best-fitting $\rm \sigma_{\rm s}$-$M_{\rm h}$ relations for all six stellar mass bins. 

We then combine the $\sigma_{\rm s}$ of galaxy sample with its corresponding $\rm \sigma_{\rm s}$-$M_{\rm h}$ relation to derive the halo mass. The errors of the halo masses are obtained by considering both the uncertainties in $\sigma_{\rm s}$ and the $M_{\rm h}$-$\sigma_{\rm s}$ relation. Both $\sigma_{\rm s}$ and the $M_{\rm h}$-$\sigma_{\rm s}$ relation are derived by using MCMC fitting, which can provide the posterior distributions for them. For galaxies in a given stellar mass bin, we can randomly generate a $\sigma_{\rm s}$ value and a $M_{\rm h}$-$\sigma_{\rm s}$ relation respectively from their posterior distributions, and predict a halo mass by combining the two. In practice, we generate 30,000 predictions of the halo mass for each subsample, and use the 16 and 84 percentiles of the mass distribution to represent the uncertainties. The halo mass estimated in this way is referred to as the weak lensing calibrated satellite kinematics (LcSK) halo mass, and is denoted also by $M_{\rm h}$.

Finally, we want to note that the uncertainties of the LcSK masses in different stellar mass bins are correlated. Therefore, if one wants to use these data points to constrain a galaxy formation model,  one should
consider the covariance among different stellar mass bins.

\begin{figure}
\includegraphics[scale=0.5]{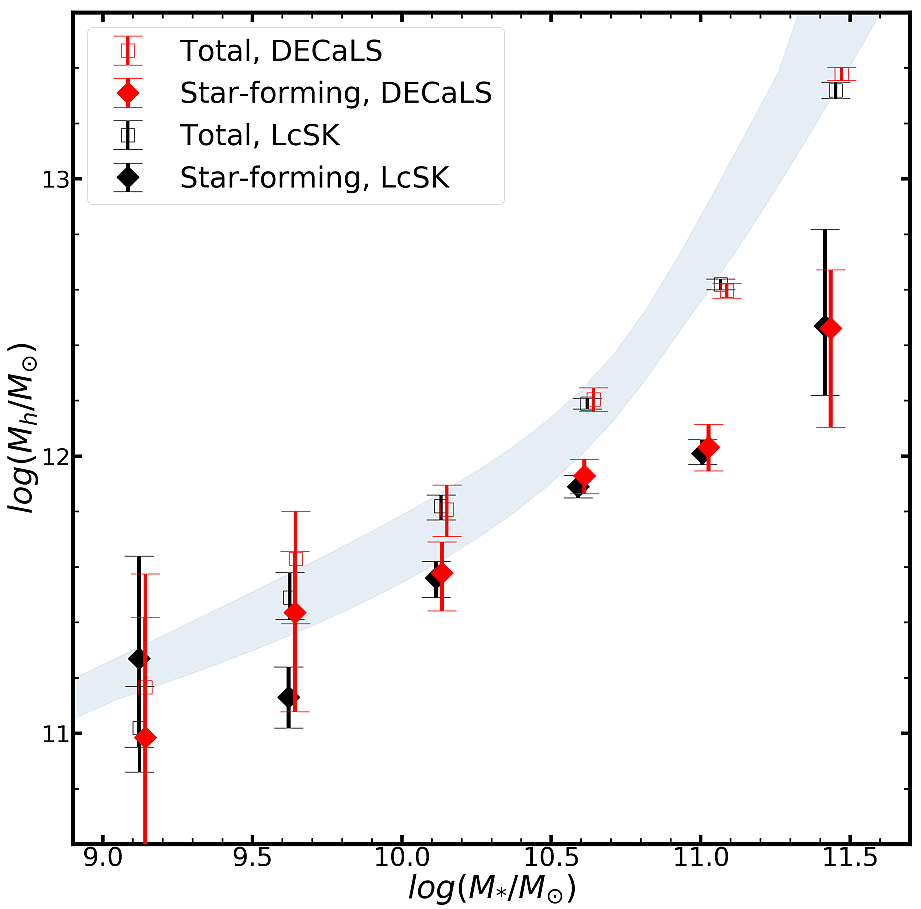}
\centering
\caption{Stellar mass - halo mass relation (SHMR).
The red symbols with error bars show the SHMR calculated by using the halo mass obtained from the fits to the stacked lensing mass profiles from the DECaLS shear catalog, 
while the black ones are based on the weak lensing calibrated satellite kinematics(LcSK) method. For clarity, we shift the black symbols towards the left slightly.
The error bars reflect the 16 and 84 percentiles of the posterior distribution.
The diamonds and squares show the results for star-forming and total populations, 
respectively. The shadow region represents the range covered by 
curves published in 
\citet{Yang2009,Moster2010,Leauthaud2012,Kravtsov2018,Behroozi2019}. 
}\label{fig_SHMR}
\end{figure}

\begin{figure*}
\includegraphics[scale=0.63]{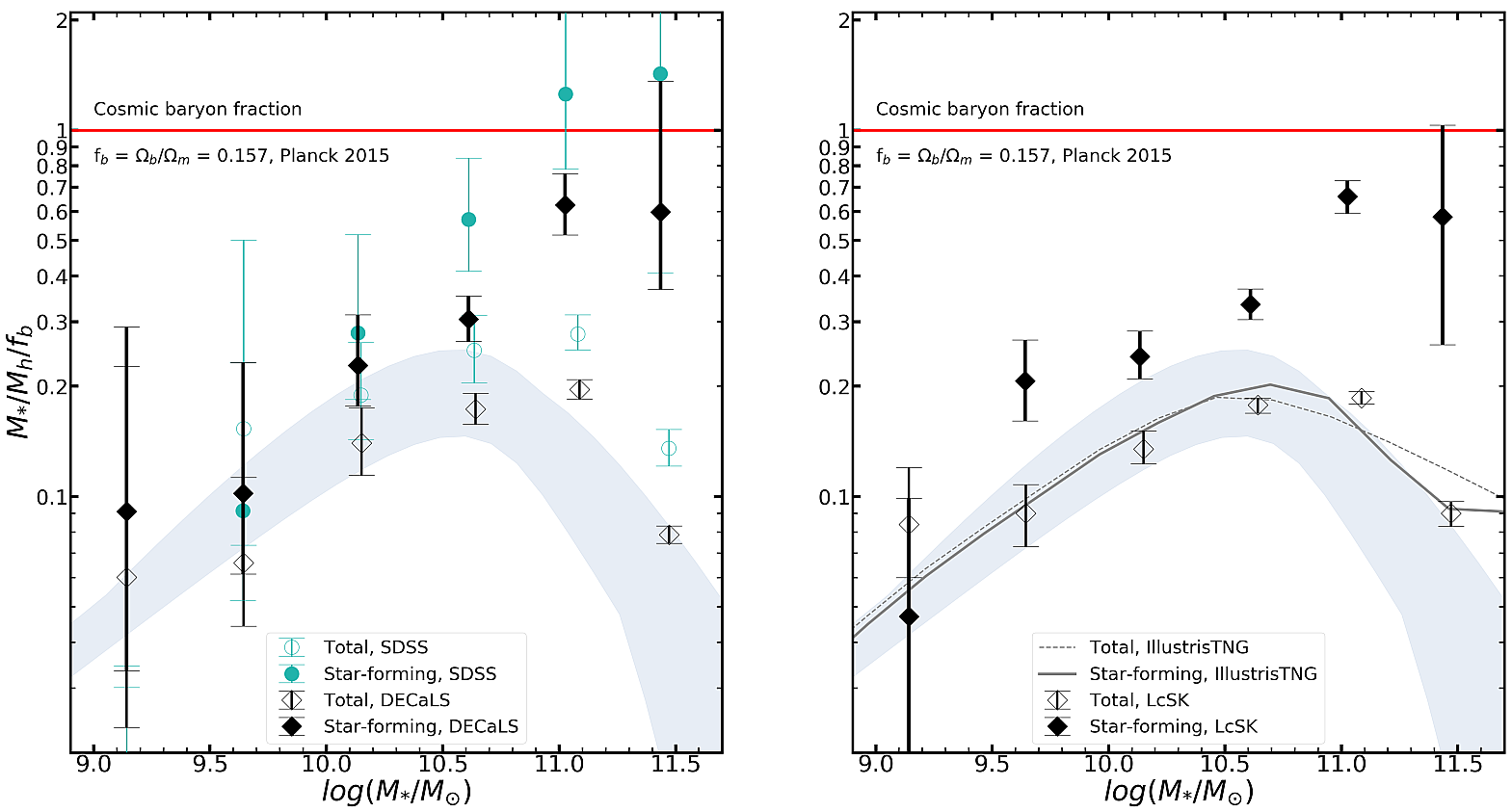}
\centering
\caption{Baryon conversion efficiency as a function of stellar mass. 
In the left panel, the symbols with error bars show the efficiency calculated 
by using the halo mass obtained from the fits to the stacked lensing mass
profiles. The black ones show the results from the DECaLS shear catalog, 
while the green ones are obtained from the SDSS shear catalog (see Appendix \ref{luo_res}).
The solid and open symbols show the results for star-forming and total populations, respectively.
Error bars indicate the 16 and 84 percentiles of the posterior distribution obtained from the MCMC fitting. 
In the right panel, the symbols with error bars show the efficiency calculated by using 
the LcSK method. 
The error bars reflect the 16 and 84 percentiles of the posterior distribution
obtained from the LcSK.
The grey solid (dashed) lines show the results for star-forming (total) galaxies 
in the hydro-simulation IllustrisTNG.
The shadow region in two panels covers the range obtained 
before, as in Figure \ref{fig_SHMR}. 
}\label{fig_eff}
\end{figure*}

\subsection{Galaxy clustering analysis}\label{sec_gcl}

Galaxy clustering can be used to measure the large scale environment 
and to infer the halo mass for a sample of selected galaxies. 
In this paper, we adopt the projected two-point cross-correlation function (2PCCF) 
to quantify the clustering of galaxies. 
We first estimate the 2PCCF using
\begin{equation}
\xi (r_{\rm p}, r_{\pi}) = \frac{N_{\rm R}}{N_{\rm D}}\frac{GD(r_{\rm p}, r_{\pi})}{GR(r_{\rm p}, r_{\pi})}-1,\\  
\end{equation}
where $N_{\rm R}$ and $N_{\rm D}$ are the galaxy numbers in the 
random and reference samples, respectively; $r_{\rm p}$ and $r_{\pi}$ 
are the separations perpendicular and parallel to the line 
of sight, respectively; $GD$ is the number of cross pairs between 
the selected galaxy sample and the reference sample and $GR$ is that 
between the selected galaxy sample and the random sample. 
To obtain the projected 2PCCF, we integrate $\xi$ along the line of sight ,  
${ w_{\rm p}(r_{\rm p}}) 
    = \int_{-\Delta s}^{\Delta s}\xi(r_{\rm p},r_{\pi})dr_{\pi}$, 
with $\Delta s=40\mpc$, sufficiently large so as to include almost all correlated pairs.
The errors on the measurements of the 2PCCF are estimated by using 100
bootstrap samples. 

 The reference sample used here is the same as that described in 
Subsection~\ref{sec_skwl}. Based on this reference sample, 
we construct a random sample in the following way
\citep[see also][for more details]{Li06a}. We generate ten duplicates for each galaxy 
in the reference sample and randomly place them in the SDSS sky coverage. 
All other properties, including stellar mass and redshift 
of the duplicate galaxies, are the same as those of the original galaxy. 
Thus, the random sample has the same survey geometry and the same 
distribution of galaxy properties as the reference sample.

The 2PCCFs of the star-forming galaxy sample cannot be directly compared 
with that of the total galaxy sample, because the two samples may have different 
redshift distributions. In order to make a fair comparison, it is necessary to 
construct a control sample that matches the star-forming sample. 
For a star-forming galaxy in a given stellar mass bin, we select,
from the total galaxy subsample in the same stellar mass bin,
$n$ galaxies whose redshift are within 0.005 from the star-forming 
galaxy in question. We choose $n=[1, 1, 1, 2, 4, 19]$ for the six 
galaxy stellar mass bins (see Section \ref{sec_sp}).
The number of control galaxies, $n$, is chosen according to 
the ratio in size between total and star-forming galaxy 
subsamples. This control sample is then used to estimate 
the 2PCCFs to compare with the corresponding star-forming galaxies.   

\section{Results}\label{sec_res}

\subsection{The stellar mass - halo mass relation}

Figure \ref{fig_SHMR} and Table \ref{table1} show the halo mass ($M_{\rm h}$) 
obtained from weak lensing (lensing mass) and from the weak lensing calibrated satellite kinematics (LcSK) method, 
separately for the total and star-forming central 
galaxies of different stellar masses. Both estimates give very similar results,
as is expected from the tight correlation between the velocity dispersion and 
the lensing mass (Figure \ref{fig_sigmamh}).
For comparison, we also show the results in the literature 
obtained using various methods, including galaxy groups, 
abundance matching, conditional luminosity function, weak lensing, 
and empirical model
\citep{Yang2009,Moster2010,Leauthaud2012,Kravtsov2018,Behroozi2019}.
As one can see from the figure, the stellar mass-halo mass relation (SHMR)
for our total galaxy sample follows closely the trend defined by 
previous results. In contrast, for a given stellar mass, the halo mass
of star-forming galaxies is lower than that of the total sample, 
and the difference becomes larger and more significant at the 
high-mass end. This is broadly consistent with previous weak-lensing and 
satellite kinematics studies that found star-forming/blue galaxies reside 
in less massive halos than quiescent/red galaxies of the same stellar mass
\citep[see e.g.][]{Mandelbaum2006, More2011, Mandelbaum2016, Lange2019, Zhang2021}, 
and with results obtained from some empirical models constrained
by observational data
\citep[e.g.][]{Rodriguez-Puebla2015ApJ...799..130R, Behroozi2019}. 

\begin{figure}
\includegraphics[scale=0.4]{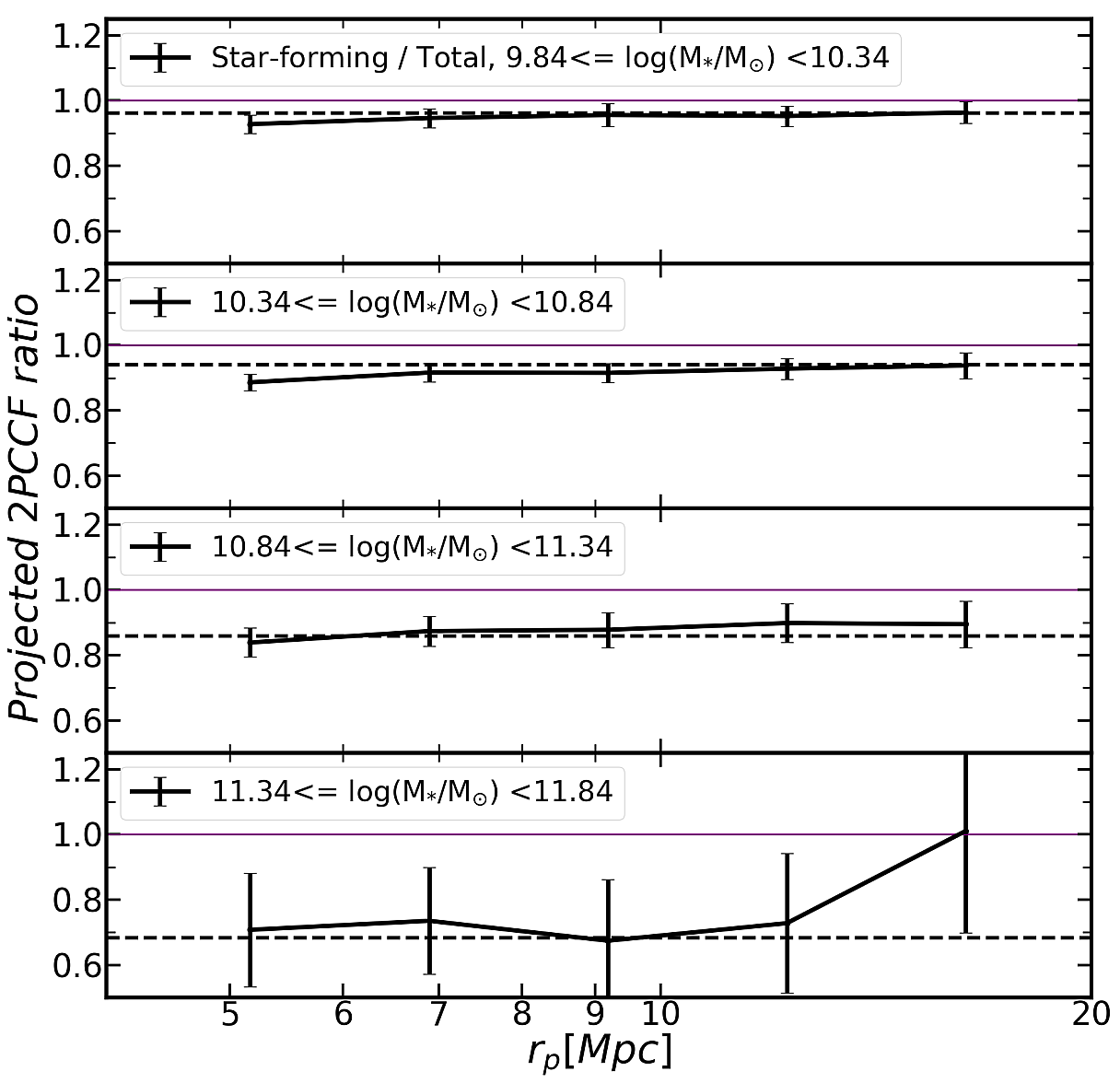}
\centering
\caption{The ratio of the 2PCCF between star-forming galaxies and 
the total control galaxies as a function of the projected distance ($r_{\rm p}$) 
in different stellar mass bins. Error bars are the standard deviation of the 2PCCF ratio
among 100 bootstrap samples. The horizontal dashed line in each panel 
indicates the theoretical ratio from a halo bias model \citep{Tinker2010} 
using the halo mass measured from weak lensing as the input. For clarity, we only 
show the results for the four highest-mass bins, where the uncertainties in 
lensing mass measurements are relatively small.}
\label{fig_2pccf}
\end{figure}

\subsection{The conversion efficiency}

Assuming that the baryon fraction within halos is equal to the cosmic fraction ($f_{\rm b}$), 
the overall conversion efficiency can be represented by 
$M_*/(f_{\rm b}M_{\rm h})$. Here $M_{\rm h}$ is the total mass of a dark matter halo, 
$M_*$ is the stellar mass of the central galaxy within the halo. In the left panel of
Figure \ref{fig_eff}, we show this conversion efficiency as a
function of stellar mass for the total galaxy sample, with the halo mass 
measured directly from the weak lensing data. The efficiency for the total galaxy 
sample peaks roughly at $M_*\sim10^{10.6}\Msun$ and decreases 
towards both lower and higher mass ends, in good agreement with previous results
\citep[e.g.][]{Wechsler2018}. 
The same panel also shows the results for star-forming central 
galaxies defined in Section \ref{sec_sp}.
The efficiency of star-forming galaxies follows roughly the trend of the total
population at the low-mass end,  but deviates from it and continues the 
increasing trend toward the massive end. 
For star-forming galaxies of $M_*\sim10^{11}\Msun$, 
the efficiency reaches a value of $0.629^{+0.142}_{-0.109}$, much higher than 
the maximum value of $\sim 0.2$ for Milky-way-like galaxies
with $M_*\sim 10^{10.6}\Msun$. At $M_*>10^{11}\Msun$, the high efficiency
seems to remain, although with a lager uncertainty.
This high efficiency indicates that the most of the baryonic gas in the 
host halos of those massive star-forming galaxies has already 
assembled into galaxies and been converted into stars. 

We can also use the halo mass obtained from the LcSK method 
to estimate the conversion efficiency
(Section \ref{sec_skwl}), and results are presented in the right panel of Figure \ref{fig_eff} 
and in Table \ref{table1}. The results are very similar to those obtained from the 
weak lensing data. For example, the efficiency for the second most massive 
star-forming sample is $0.66^{+0.069}_{-0.067}$.  However, the LcSK method
significantly reduces the uncertainty of the efficiency in most of the stellar mass bins. 

As mentioned above, many previous studies found that star-forming or blue 
galaxies tend to reside in smaller halos than quenched or red galaxies of 
the same stellar mass\cite[e.g.][]{More2011, Hudson2015, Mandelbaum2016, Lange2019, Behroozi2019, Posti2019, Bilicki2021}. 
This is consistent with our results that the conversion efficiency is higher for 
star-forming galaxies than for quenched galaxies. However, in comparison with our results, the implied conversion efficiency is 
significantly lower and has larger uncertainties in many of the earlier studies. 
For example, \cite{Mandelbaum2006} and \cite{Mandelbaum2016} found a peak 
efficiency of $0.35^{+0.92}_{-0.17}$ for late-type galaxies and $0.351^{+0.165}_{-0.089}$ 
for blue galaxies, respectively.  \cite{Dutton2011} found that the mean peak efficiency 
for their late-type galaxies is around 0.3. \cite{Rodriguez-Puebla2015ApJ...799..130R} 
found that the $\rm M_*$-to-$\rm M_h$ ratio of their blue centrals has a peak value of  
0.051, corresponding to a conversion efficiency of $0.325$ assuming $f_{\rm b}=0.157$. 
These results appear to be in conflict with ours, and we will come back to this issue 
later. Recently, \cite{Posti2019} modeled the rotation curves of local disc 
galaxies and inferred the halo mass of individual galaxies. They found that the mean
conversion efficiency for about 20 massive disc galaxies is about 0.5 
but with large uncertainties. Our results are in broad agreement with theirs.
However, our results are obtained from a large sample of 22,099 star-forming 
galaxies, indicating that the high conversion efficiency is a common
property for massive star-forming galaxies.

\subsection{Tests of uncertainties}\label{sec_tests}

A number of factors may affect the estimates of the halo mass. 
As a test, we repeat our analysis using an independent weak-lensing 
shear catalog obtained by \citet{Luo2017ApJ} from the Sloan Digital Sky Survey 
(SDSS) DR7 \citep{Abazajian-09} imaging data using a totally different 
method (see Appendix \ref{luo_res}). As shown in Figure \ref{fig_eff}, the  
results obtained from the SDSS data agree well with those from 
the DECaLS data. However, the SDSS results have much larger 
uncertainties because of the shallower imaging data used to measure 
the weak lensing shear. 

Galaxy clustering provides another test because of its 
dependence on halo mass \citep{Mo1996,Tinker2010}. 
Figure \ref{fig_2pccf} shows the ratio of the 2PCCF, in the range of 
$5\Mpc<r_{\rm p}<20\Mpc$, between star-forming galaxies
and total population in four high-mass bins
(see Section \ref{sec_gcl} for the method to estimate the 2PCCF). 
The horizontal dashed line in each panel shows the theoretical prediction 
using the linear halo bias model \citep{Tinker2010} 
and the halo mass was derived from the DECaLS weak lensing 
measurements. As one can see, the model predictions agree very well 
with the observational results on large scales where the linear halo bias model is valid. 
However halo bias may also depends on halo assembly history, in addition to halo 
mass \citep[e.g.][]{Gao2005, WangH2007}. 
If there is a correlation between the properties of the 
central galaxy in a halo and the assembly of the halo, 
then the halo bias model used here may not be an accurate
description for star-forming and quenched centrals. 
Unfortunately, this potential correlation is not understood 
well enough to quantify the effect which may produces on our results. 

\begin{figure}
    \centering
    \includegraphics[scale=0.395]{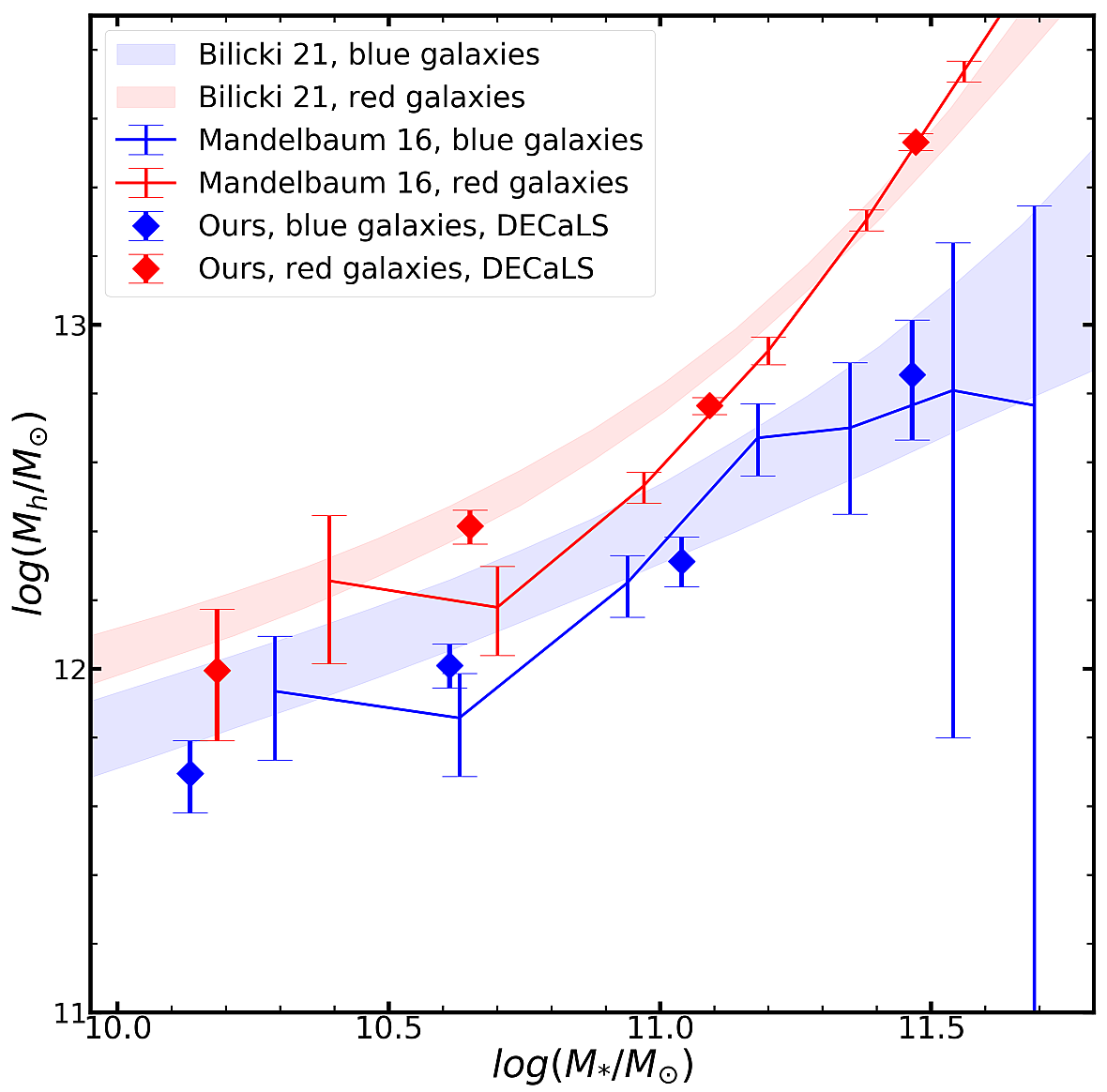}
    \caption{Comparison of our SHMR with \cite{Mandelbaum2016} and \cite{Bilicki2021}. The diamonds, the lines with error bars and the shaded regions are the SHMR from our results, \cite{Mandelbaum2016} and \cite{Bilicki2021}, respectively. The solid symbols show the results calculated by using DECaLS shear catalog. The red and blue color represent the red and blue galaxy samples, respectively.}
    \label{fig_redblue}
\end{figure}

As discussed above, the conversion efficiency obtained here for massive 
star-forming galaxies appears to be higher than that obtained in 
some previous investigations. One cause of the discrepancy may be that 
galaxy samples used in these investigations are different from ours. 
For example, \cite{Mandelbaum2016} and \cite{Bilicki2021} split galaxy samples according to galaxy 
color, instead of the SFR used here. As a test,  we have repeated 
our analysis by separating red and blue galaxies 
according to equation (1) in \cite{vandenBosch2008}, a way  
similar to that in \cite{Mandelbaum2016}. The SHMRs 
obtained from the red and blue populations are shown in Figure \ref{fig_redblue}, 
in comparison with the results taken from the table B1 of \cite{Mandelbaum2016} 
\footnote{To be consistent, the stellar masses are adopted from the MPA/JHU catalog, and their halo masses are uncorrected; see the paper for the correction.} 
and from \cite{Bilicki2021}. 
As one can see, our results are in 
good agreement with theirs, which provides additional support to 
the reliability of our mass estimates. In particular, blue galaxies of $M_*\sim10^{11}\Msun$ 
have an average halo mass of $12.31^{+0.07}_{-0.07}$ from \cite{Mandelbaum2016} and $12.39^{+0.15}_{-0.08}$ from \cite{Bilicki2021}, corresponding to a conversion efficiency of
 $0.338^{+0.062}_{-0.052}$  and $0.262^{+0.053}_{-0.077}$, respectively, 
 much lower than that for star-forming galaxies
of the same stellar mass (Fig. \ref{fig_eff} and Tab. \ref{table1}). 
This suggests that the calculated efficiencies are sensitive to the sample selection, 
and that the discrepancy between our results (samples selected by the star formation rate) 
and their results (samples selected by galaxy colors) is entirely due to the difference 
in the sample selection. 

Other potential problems may also exist in the halo mass estimation.   
If a fraction of the selected galaxies are satellites, instead of centrals, 
a systematic bias can be introduced, as our modeling of the lensing 
measurements assumes that all galaxies are centrals. 
Contamination by satellites is expected to lead to an overestimation of the 
halo mass \citep{Mandelbaum2006,Mandelbaum2016}, and so the conversion
efficiency may be underestimated in our results.
To check the impact of this effect, we make an analysis by 
using an additional selection criterion to reduce the contamination
of satellites in our central galaxy samples. Specifically,  
we require that a central galaxy be the most massive one  
among all its neighbors that have projected 
distances less than $1\Mpc$, and line-of-sight velocity 
differences smaller than $1000\kms$, relative to the galaxy in question.  
This leads to a sample of 4,900 star-forming galaxies in the second 
most massive bin. The halo mass obtained from lensing for this 
sample is $\log M_{\rm h}/\Msun=12.06^{+0.16}_{-0.21}$, very close 
to the value obtained above, demonstrating that the effect of 
satellite contamination is negligible in our results. 

Another bias in the halo mass estimate may arise because 
halos of galaxies in a given sample can span a large range in mass. 
The stacked lensing signal around the sample 
galaxies is, therefore, an average of many different halo profiles, and 
using a single NFW profile to fit the average profile may introduce a bias. 
Detailed tests by \cite{Mandelbaum2016} using 
a mock catalog constructed from a semi-analytic galaxy formation model 
suggest that the best-fitting mass underestimates the mean halo mass by 
about $10\%$, quite independent of the stellar mass and galaxy color. 
At $M_*\sim10^{11}\Msun$, the actual mean halo mass may be about 1.14 
times the best fitting halo mass. Taking this correction into account, 
the conversion efficiency for star-forming galaxies in this mass range 
changes to $0.629/1.14=0.552$. One should keep in mind that the correction factor may depends on the used galaxy formation model. 
As we will see below, our results are not well reproduced 
by the current galaxy formation models. 

\begin{figure}
    \centering
    \includegraphics[scale=0.38]{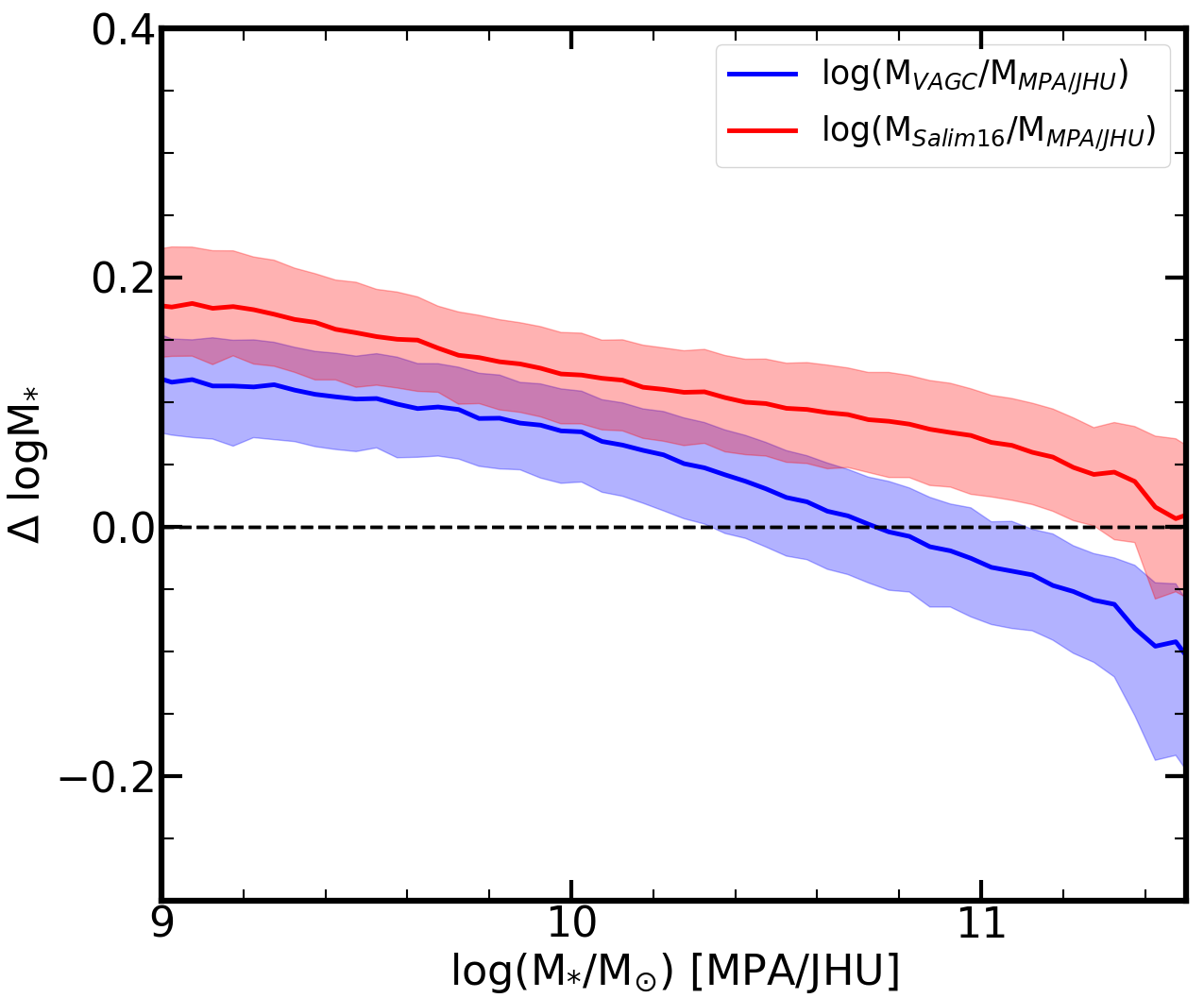}
    \caption{Comparison of the MPA/JHU stellar mass with the stellar masses from NYU-VGAC (blue) and \cite{Salim2016ApJS} (red) for central star-forming galaxies.  The solid lines show the median value of the residual, while the shaded regions show the 25 and 75 percentiles of the residual distribution.}
    \label{fig_comms}
\end{figure}

Yet another important source of uncertainty comes from the estimate of the stellar mass. 
The statistical uncertainty of the stellar mass is usually 0.3 dex 
\citep[e.g.][]{Kauffmann2003b}. Our galaxy sample
is large enough, so that the statistical error in the mean stellar mass 
is small. However, the systematic uncertainties, such as those produced by the 
adopted initial mass function, the star formation history, the stellar library, and 
the dust attenuation, may not be negligible 
\citep[see][for a brief introduction]{Moustakas2013}. 
It is in general difficult to evaluate such systematic uncertainties 
within a specific model of the stellar population.  
One common practice is to make a consistency check by comparing the stellar masses 
of the same object measured with different techniques and/or from different data. 
The MPA masses used here have been used in many comparison studies in the literature.
For example, \cite{Moustakas2013} showed that the MPA masses are
in excellent agreement with theirs based on the fits to SEDs in 12 
UV, optical and infrared bands. As a check, here we make a similar  
comparison for central star-forming galaxies, as they are the most relevant 
to our results. In Figure \ref{fig_comms}, we compare the MPA masses 
with those given in NYU-VAGC and those obtained by \cite{Salim2016ApJS}. 
Different approaches and/or data were used to derive stellar masses
in these three databases. As one can see, there are some 
systematic differences among the three masses. At $M_*\sim10^{11}\Msun$, 
the difference between the MPA mass and the other two is about 0.05 dex, 
and MPA mass appears to lie between the two masses. 
Thus, our results are robust as long as the systematic bias in the stellar mass is not much larger than the difference among the three mass estimates compared here. 

\section{Discussion}\label{sec_dis}

\begin{figure}
\includegraphics[scale=0.38]{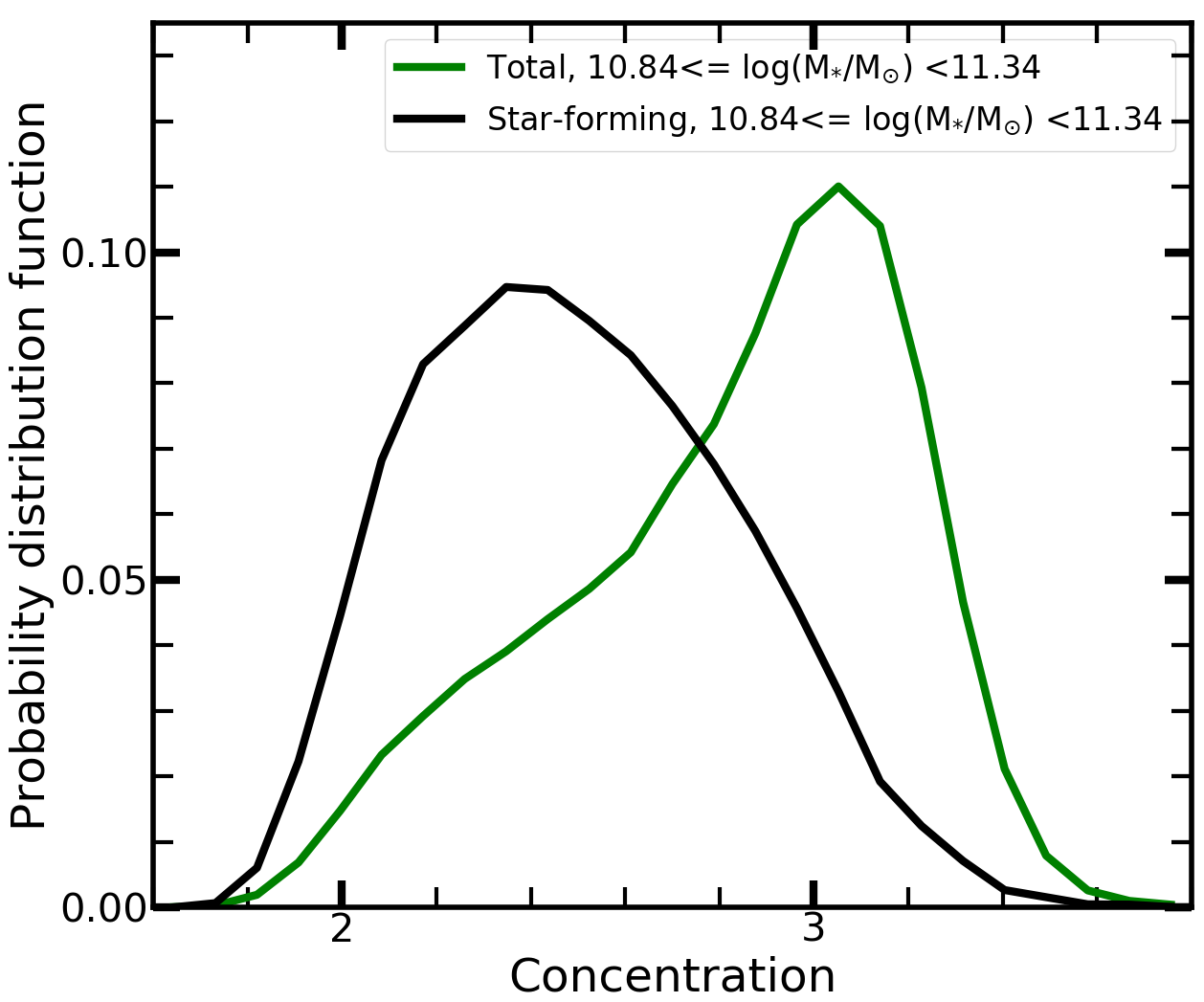}
\centering
\caption{The probability distributions of galaxy concentration (C$=R_{90}/R_{50}$) for total galaxy sample (green line) and massive star-forming galaxy sample (black line) of $M_*\sim 10^{11}\Msun$.}\label{fig_gp}
\end{figure}

One possibility for the large $M_*$ to $M_{\rm h}$ ratio reported here is that their host
halos are splashback halos \citep[e.g.][]{Ludlow2009}, which have ever entered the virial radii of 
more massive halos, and were severely stripped by the tidal field. However, splashback halos are usually much more strongly clustered than other halos 
of the same mass \citep{Wang2009}, because they are spatially 
close to massive halos. This is clearly inconsistent with the fact that 
the massive star-forming galaxies are less clustered than the total population 
of the same mass (Figure \ref{fig_2pccf}). Moreover, environmental effects may 
quench the star formation in splashback halos, and so their galaxies 
are not expected to be star forming. Thus, this possibility can be ruled out.

Another possibility is that some processes that are supposed to prevent the 
growth of massive galaxies did not operate on these galaxies in the past.
Figure \ref{fig_gp} shows the distributions of galaxy concentration
for star-forming and total galaxies with $M_*\sim 10^{11}\Msun$. 
One can see that star-forming galaxies have much 
smaller concentrations than the total galaxies of the same mass.
AGN activities, which are thought to be capable of quenching star formation, 
are expected to be more prominent in more concentrated galaxies
\citep[e.g.][]{Kauffmann2003, Zhang2021}. 
Previous theoretical and observational studies found that the efficiency 
of star-forming galaxies may be suppressed by several feedback processes, 
such as supernova and AGN feedback in the low and high stellar mass ends, respectively. 
Our results suggest that AGN feedback must be inefficient in suppressing 
cold gas acquisition and star formation in massive star-forming galaxies.
AGN feedback may still be effective in quenching the star formation in
other massive galaxies of $M_*\sim10^{11}\Msun$, and thus to produce a much lower conversion efficiency 
in them. As shown in Table \ref{table1}, the number of the
 massive star-forming galaxies ($M_*\sim10^{11}\Msun$) is much smaller than 
that of the total galaxy population of the same mass. Thus this absence of effective 
AGN feedback only applies to a relatively small fraction of the total galaxy population.

These massive star-forming galaxies have already converted 
more than sixty percent of their halo gas into stars.
Based on CO and HI observations, \cite{Saintonge2016} 
found that these galaxies on average contain about $\sim10^{10}\Msun$ in cold gas.
Thus the total baryon mass in these galaxies is more than seventy percent
of the total baryons in their halos. This leaves less than thirty percent of 
the baryons in the circumgalactic medium (CGM). 
Thus, observations of the CGM may provide an independent way 
to check our results.
CGM can be probed in 
a number of ways, such as quasar absorption line systems
\citep[e.g.][]{Tumlinson2017}, extended X-ray emission \citep[e.g.][]{Anderson2013ApJ}, 
and SZ effect \citep[e.g.][]{Lim2018}. Many studies found evidence 
for the existence of CGM in galaxies with a wide range of stellar mass
\citep[e.g.][]{Tumlinson2017}. However, a detailed comparison with our 
results is not straightforward. As shown in Table \ref{table1}, star-forming 
galaxies of $10^{11}\Msun$ make only about one-fifth of the total population of 
the same mass. It is thus inappropriate to compare our results 
directly with those that did not separate the star-forming galaxies from the 
total population. For low-ionization line systems observed in  
star forming galaxies, such as MgII, the absorbing gas is believed to be 
associated with outflow \citep[e.g.][]{Lan2018ApJ}. The total amount of gas 
involved may not be large, which is consistent with our results.  

The mean density of the CGM around massive 
star-forming galaxies is expected to be less than that around other galaxies. 
Therefore, the timescale for gas cooling, which is 
inversely proportional to the gas density, is much longer.
The ability of massive star-forming galaxies to 
acquire additional gas to maintain a
high star formation rate should be significantly suppressed.
Our results suggest that the well-known flattening of the SFR$-M_*$ relation 
for star-forming galaxies at the massive end \citep{Whitaker2014, Saintonge2017} is 
caused by the decrease of gas supply. It is consistent
with the analysis based on atomic and molecular gas within galaxies\citep{Saintonge2016}. 

\begin{figure}
    \centering
    \includegraphics[scale=0.38]{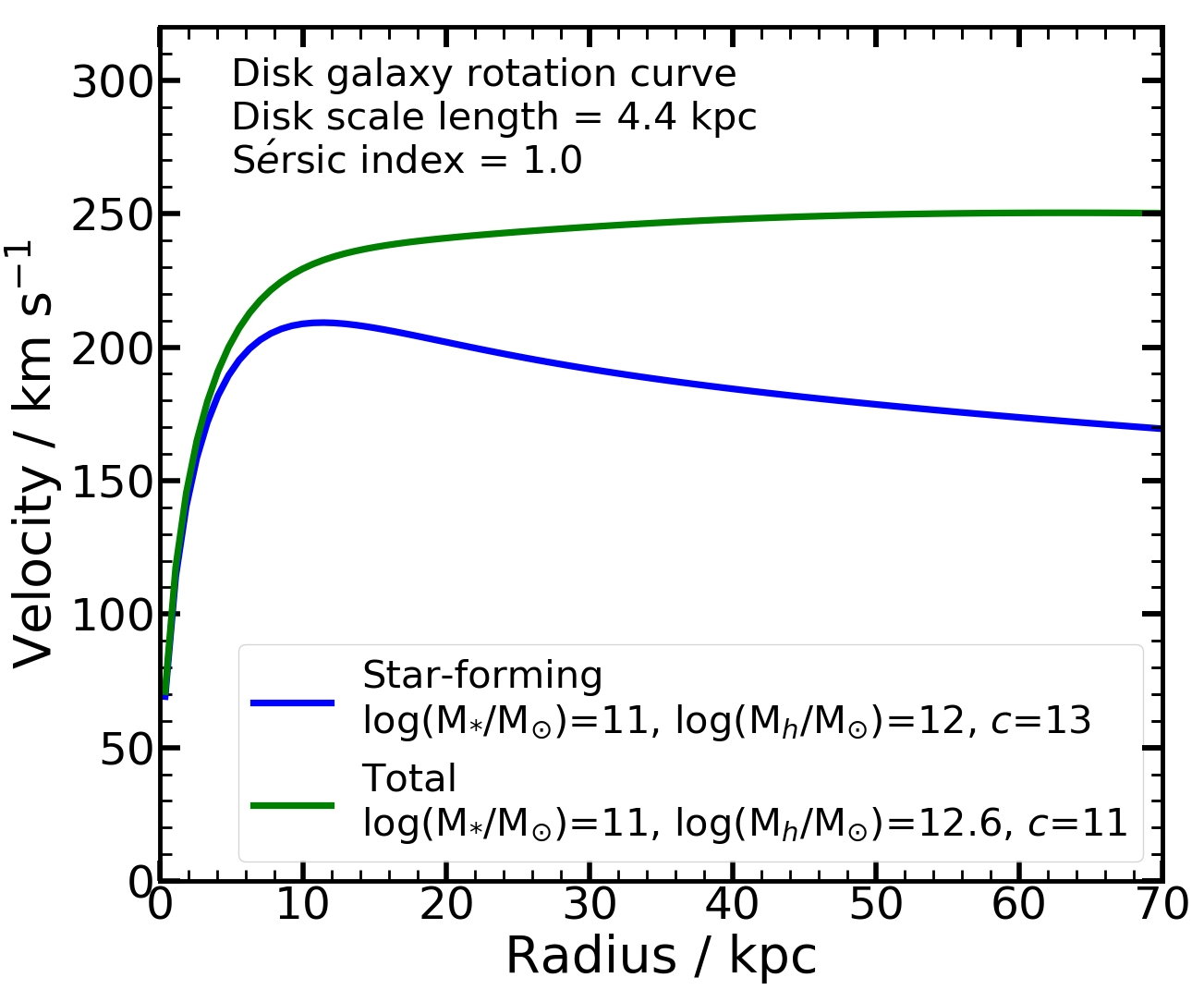}
    \caption{The figure shows rotation curves for two disk galaxies with the stellar mass of $10^{11}\Msun$ but with the halo mass of $10^{12}\Msun$ (star-forming, blue) and $10^{12.6}\Msun$ (total, green), respectively. The two disc galaxies  have the same stellar mass exponential profiles with disk scale length of $4.4\Kpc$ and S$\acute{e}$rsic index of $n=1$. Dark matter halos are assumed to be NFW profile with concentration-mass relation from \cite{Zhao2009ApJ}, $c=13$ for the blue line and $c=11$ for the green line.}
    \label{fig_rc}
\end{figure}

The high galaxy-to-halo mass ratio for massive star-forming galaxies 
may have important implications for their rotation curves.
To demonstrate this, we select galaxies with $M_*\sim10^{11}\Msun$ 
and S$\acute{e}$rsic index of $n\sim 1$ from the SDSS galaxy sample. 
These massive disc galaxies have a wide distribution of S$\acute{e}$rsic $r_0$
(which is equal to the disk scale length for $n=1$),
ranging from 2.5 to $6.3\Kpc$, where both $n$ and $r_0$
are taken from the NYU-VAGC catalog \citep{Blanton-05b}. 
To derive the rotation curve, we assume an exponential mass profile with a typical 
disk scale length of $4.4\Kpc$ for a massive disc galaxy and a NFW profile 
for its host dark matter halo. 
In Figure \ref{fig_rc}, the blue line 
shows the rotation curve for a halo mass of $M_{\rm h}=10^{12}\Msun$ (with concentration
$c=13$), about the average value for star-forming galaxies with $M_*\sim10^{11}\Msun$.
For comparison, we also show the rotation curve for a disc galaxy residing in a halo of 
$10^{12.6}\Msun$ ($c=11$), as expected from our total galaxy sample 
with $M_*\sim10^{11}\Msun$ (see Figure \ref{fig_SHMR}). 
For $M_h=10^{12.6}\Msun$, the predicted rotation curve is 
quite flat at large radius, as is observed for many disk galaxies.  
In contrast, for $M_{\rm h}=10^{12}\Msun$, the rotation curve reaches a peak 
at $\sim10\Kpc$ and gradually decreases at larger radius. 
This type of rotation curve is not common for the general disk 
population, but has been found for some local galaxies \citep[e.g.][]{Corbelli2010A&A, Posti2019}. 
Modeling their rotation curves show that these galaxies indeed have 
a high conversion efficiency \citep{Posti2019}. Our results 
show that this type of rotation curve should be expected for 
massive star-forming disks.   

Finally, we examine whether current galaxy formation models
can reproduce our results. In the right panel of Figure \ref{fig_eff}, 
we show the results obtained from the Illustris-TNG simulation \citep{Pillepich2018}, 
which implements a series of baryonic physics, such as AGN feedback, 
to suppress star formation in massive galaxies (see Appendix \ref{TNG_res} for a brief 
description of the simulation). 
As can be seen from the figure, the simulations result for the total sample
has a peak around a stellar mass between $10^{10.5}$ and $10^{11}\Msun$,
consistent with the observational results. However,  the simulation 
fails to reproduce the high conversion efficiency for the observed massive star-forming galaxies. Indeed, massive star-forming galaxies in the simulation follow closely with the total population over almost the entire stellar mass range. 
It is likely that the AGN feedback implemented in the simulation is 
too strong for these galaxies. We have also examined the Eagle simulation \citep{Schaye2015}, 
and found also that it cannot reproduce the high conversion efficiency 
observed for massive star-forming galaxies.
Thus, our finding presents a challenging problem for current simulations
in their modeling of feedback and star formation.

\section{Summary}\label{sec_sum}

Based on the shear catalog of DECaLS imaging data, we derive the
halo mass of central galaxies selected from the SDSS. 
We develop a weak lensing calibrated satellite kinematics method and 
to improve the halo mass measurements. We then obtain the efficiency 
for converting baryons into stars within halos, defined as $M_*/M_{\rm h}/f_{\rm b}$, 
for both the total galaxy population and galaxies in the
star-forming main sequence. Our main results are summarized as follows.

\begin{itemize}
    \item The stellar mass-halo mass relation for the total galaxy population we obtained is in good agreement with previous studies. The conversion efficiency peaks around Milky-Way-like galaxies and declines towards both lower and higher stellar mass ends.
    \item The conversion efficiency of star-forming galaxies increases monotonically with stellar mass, and reaches a value of more than sixty percents at $M_*\gtsima10^{11}\Msun$. Thus, these galaxies have converted most of their halo gas into stars.
    \item Our tests show that the measurements of the halo mass are consistent with the results obtained from the SDSS shear catalog and from galaxy clustering.
    \item Massive star-forming galaxies are expected to have rotation curves that are peaked at about two disk scalelengths and decline at larger distances, quite different from the flat rotation curves commonly observed for the general disk population.
    \item The high conversion efficiency observed for massive star-forming galaxies is not reproduced by current cosmological gas simulations.
    \item We have tested a number of systematic effects that may affect our results and found that none of them can change our conclusions significantly. 
\end{itemize}

Our finding has important implications for understanding galaxy formation and star formation 
quenching. The high conversion efficiency observed for massive star-forming galaxies 
indicates that AGN feedback may not have played an important role in affecting the 
conversion of gas into stars in these particular galaxies. The fact that current cosmological hydrodynamic simulations 
cannot reproduce such a high conversion efficiency indicates that our current 
understanding of feedback is still incomplete, at least for massive star-forming 
galaxies. Clearly, the observational results presented here will provide 
an important constraint on modeling feedback processes in galaxy formation, and we will come back to this 
in a future paper.

\section*{Acknowledgments}
We thank the referee for useful comments. This work is supported by the National Key R\&D Program of China (grant No. 2018YFA0404503; grant No. 2018YFA0404504), the National Natural Science Foundation of China (NSFC, Nos.  11733004, 12192224, 11890693, 11421303, 11890691, 11621303, 11833005 and 11890692), and the Fundamental Research Funds for the Central Universities. The authors gratefully acknowledge the support of Cyrus Chun Ying Tang Foundations. We acknowledge the science research grants from the China Manned Space Project with No.
CMS-CSST-2021-A03. The work is also supported by the Supercomputer Center of University of Science and Technology of China.

The Legacy Surveys consist of three individual and complementary projects: the Dark Energy Camera Legacy Survey (DECaLS; Proposal ID \#2014B-0404; PIs: David Schlegel and Arjun Dey), the Beijing-Arizona Sky Survey (BASS; NOAO Prop. ID \#2015A-0801; PIs: Zhou Xu and Xiaohui Fan), and the Mayall z-band Legacy Survey (MzLS; Prop. ID \#2016A-0453; PI: Arjun Dey). DECaLS, BASS and MzLS together include data obtained, respectively, at the Blanco telescope, Cerro Tololo Inter-American Observatory, NSF’s NOIRLab; the Bok telescope, Steward Observatory, University of Arizona; and the Mayall telescope, Kitt Peak National Observatory, NOIRLab. The Legacy Surveys project is honored to be permitted to conduct astronomical research on Iolkam Du’ag (Kitt Peak), a mountain with particular significance to the Tohono O’odham Nation.

NOIRLab is operated by the Association of Universities for Research in Astronomy (AURA) under a cooperative agreement with the National Science Foundation.

This project used data obtained with the Dark Energy Camera (DECam), which was constructed by the Dark Energy Survey (DES) collaboration. Funding for the DES Projects has been provided by the U.S. Department of Energy, the U.S. National Science Foundation, the Ministry of Science and Education of Spain, the Science and Technology Facilities Council of the United Kingdom, the Higher Education Funding Council for England, the National Center for Supercomputing Applications at the University of Illinois at Urbana-Champaign, the Kavli Institute of Cosmological Physics at the University of Chicago, Center for Cosmology and Astro-Particle Physics at the Ohio State University, the Mitchell Institute for Fundamental Physics and Astronomy at Texas A\&M University, Financiadora de Estudos e Projetos, Fundacao Carlos Chagas Filho de Amparo, Financiadora de Estudos e Projetos, Fundacao Carlos Chagas Filho de Amparo a Pesquisa do Estado do Rio de Janeiro, Conselho Nacional de Desenvolvimento Cientifico e Tecnologico and the Ministerio da Ciencia, Tecnologia e Inovacao, the Deutsche Forschungsgemeinschaft and the Collaborating Institutions in the Dark Energy Survey. The Collaborating Institutions are Argonne National Laboratory, the University of California at Santa Cruz, the University of Cambridge, Centro de Investigaciones Energeticas, Medioambientales y Tecnologicas-Madrid, the University of Chicago, University College London, the DES-Brazil Consortium, the University of Edinburgh, the Eidgenossische Technische Hochschule (ETH) Zurich, Fermi National Accelerator Laboratory, the University of Illinois at Urbana-Champaign, the Institut de Ciencies de l’Espai (IEEC/CSIC), the Institut de Fisica d’Altes Energies, Lawrence Berkeley National Laboratory, the Ludwig Maximilians Universitat Munchen and the associated Excellence Cluster Universe, the University of Michigan, NSF’s NOIRLab, the University of Nottingham, the Ohio State University, the University of Pennsylvania, the University of Portsmouth, SLAC National Accelerator Laboratory, Stanford University, the University of Sussex, and Texas A\&M University.

BASS is a key project of the Telescope Access Program (TAP), which has been funded by the National Astronomical Observatories of China, the Chinese Academy of Sciences (the Strategic Priority Research Program “The Emergence of Cosmological Structures” Grant \# XDB09000000), and the Special Fund for Astronomy from the Ministry of Finance. The BASS is also supported by the External Cooperation Program of Chinese Academy of Sciences (Grant \# 114A11KYSB20160057), and Chinese National Natural Science Foundation (Grant \# 11433005).

The Legacy Survey team makes use of data products from the Near-Earth Object Wide-field Infrared Survey Explorer (NEOWISE), which is a project of the Jet Propulsion Laboratory/California Institute of Technology. NEOWISE is funded by the National Aeronautics and Space Administration.

The Legacy Surveys imaging of the DESI footprint is supported by the Director, Office of Science, Office of High Energy Physics of the U.S. Department of Energy under Contract No. DE-AC02-05CH1123, by the National Energy Research Scientific Computing Center, a DOE Office of Science User Facility under the same contract; and by the U.S. National Science Foundation, Division of Astronomical Sciences under Contract No. AST-0950945 to NOAO.

The Photometric Redshifts for the Legacy Surveys (PRLS) catalog used in this paper was produced thanks to funding from the U.S. Department of Energy Office of Science, Office of High Energy Physics via grant DE-SC0007914.

\bibliography{ref.bib}
\bibliographystyle{aa}
\clearpage

\begin{appendix}

\section{Testing the impact of halo mass uncertainties on $\sigma_{\rm s}$}\label{Mh_uncertainty}

\begin{figure}[H]
\includegraphics[scale=1.]{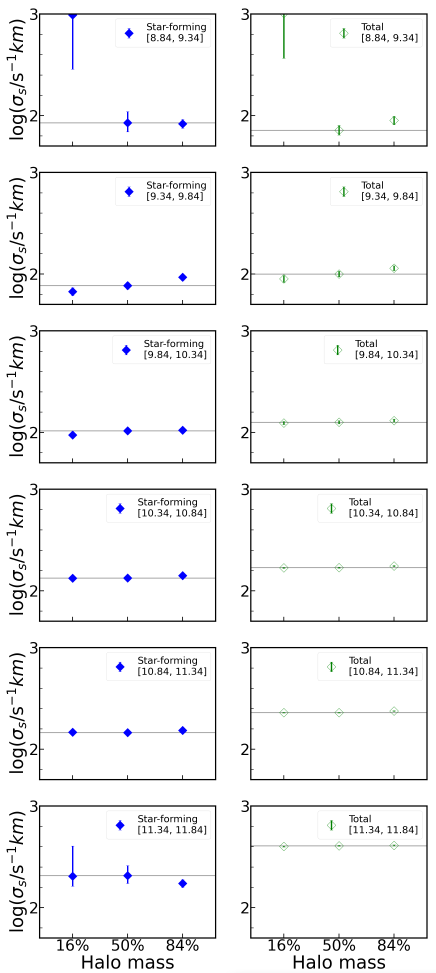}
\centering
\caption{Comparison of the satellite velocity dispersion obtained using three 
different halo masses in the estimates of $r_{\rm vir}$ and $v_{\rm vir}$. 
The left column shows the results for star-forming samples, and the right column for 
total samples. Different rows represent different stellar mass bins. 
In each panel, the three data points are for halo masses that correspond to 
16$\%$, 50$\%$ and 84$\%$ percentage points of the mass distribution,  
respectively, as labeled in the horizontal axis.}
\label{fig_delta_sigma}
\end{figure}
\FloatBarrier

We use the halo mass to determine $r_{\rm vir}$ and $v_{\rm vir}$ 
that are used to identify satellite candidates around centrals. The satellite velocity 
dispersion can thus be affected by the uncertainties in the halo mass obtained from 
weak lensing. To test this, we adopt two additional halo masses to derive the velocity dispersion. 
These two halo masses correspond to the 16 and 84 percentiles of the posterior distribution 
obtained from the MCMC fitting to the stacked lensing profiles, and may be 
considered as the lower and upper halo mass limits, respectively.  
Figure \ref{fig_delta_sigma} compares the $\sigma_{\rm s}$ obtained from the 
two halo mass limits and the medium halo mass adopted in the main text.  
The results show that, for most of the stellar mass bins where the sample sizes are 
sufficiently large, the value of $\sigma_{\rm s}$ is not affected significantly 
by the halo mass uncertainty. For some of the stellar mass bins where the sample sizes are small, such as the lowest stellar mass bin (see Table \ref{table1}), 
the halo mass uncertainty is large and may have a significant impact on 
the estimate of $\sigma_{\rm s}$.

\section{The $M_{\rm h}$-$\sigma_{\rm s}$ relations for different stellar mass bins}\label{new_Mh_sigma_relation}

To avoid the correlation between the uncertainties in $\sigma_{\rm s}$ 
and in the $M_{\rm h}$-$\sigma_{\rm s}$ relation,
we derive the $M_{\rm h}$-$\sigma_{\rm s}$ relation using a slightly different method.
For a given galaxy sample in a stellar mass bin, we use the complementary sample 
consisting of all galaxies in other stellar mass bins to fit the $M_{\rm h}$-$\sigma_{\rm s}$ relation. 
The relations so obtained for different stellar mass bins 
are shown in Table \ref{tab_new_sigmamh}. As one can see, these
relations are consistent with each other and Eq. \ref{eq_mhsigma} within error bars.
It is clear that the $\sigma_{\rm s}$ of a galaxy sample in a stellar mass bin is quite 
independent of the complementary sample used for the calibration.
Thus, we can use the value of the $\sigma_{\rm s}$ for the galaxy sample, 
in combination with the corresponding $M_{\rm h}$-$\sigma_{\rm s}$ relation, to derive the halo mass and conversion efficiency for the sample. The results are shown in Table \ref{table1}. 

\begin{table}[H]
\caption{Halo mass-satellite velocity dispersion relations.}\label{tab_new_sigmamh}
\centering
\begin{threeparttable}

\resizebox{.95\columnwidth}{!}{
\begin{tabular}{c c}
\hline
$\log M_*$ range\tnote{(a)} & $\sigma_s$-$M_{\rm h}$ relation\tnote{(b)} \\
\hline
[8.84, 9.34]   & $\log(\rm M_{\rm h}/M_{\odot})$ =(3.08$\pm$0.05)$\log(\sigma_{s}/\rm s^{-1}$ km)+(5.35$\pm$0.11).\\\relax
[9.34, 9.84]   & $\log(\rm M_{\rm h}/M_{\odot})$ =(3.09$\pm$0.05)$\log(\sigma_{s}/\rm s^{-1}$ km)+(5.32$\pm$0.12).\\\relax
[9.84, 10.34]  & $\log(\rm M_{\rm h}/M_{\odot})$ =(3.07$\pm$0.05)$\log(\sigma_{s}/\rm s^{-1}$ km)+(5.37$\pm$0.12).\\\relax
[10.34, 10.84] & $\log(\rm M_{\rm h}/M_{\odot})$ =(3.08$\pm$0.05)$\log(\sigma_{s}/\rm s^{-1}$ km)+(5.34$\pm$0.13).\\\relax
[10.84, 11.34] & $\log(\rm M_{\rm h}/M_{\odot})$ =(3.05$\pm$0.05)$\log(\sigma_{s}/\rm s^{-1}$ km)+(5.42$\pm$0.11).\\\relax
[11.34, 11.84] & $\log(\rm M_{\rm h}/M_{\odot})$ =(2.92$\pm$0.1)$\log(\sigma_{s}/\rm s^{-1}$ km)+(5.71$\pm$0.22).\\
\hline
\end{tabular}}

\begin{tablenotes}
\footnotesize
\item[(a)] The stellar mass bins.
\item[(b)] The corresponding $\sigma_s$-$M_{\rm h}$ relations.
\end{tablenotes}
\end{threeparttable}

\end{table}
\FloatBarrier

\section{Measurements with SDSS shear catalog}\label{luo_res}
\begin{figure*}
\includegraphics[scale=0.8]{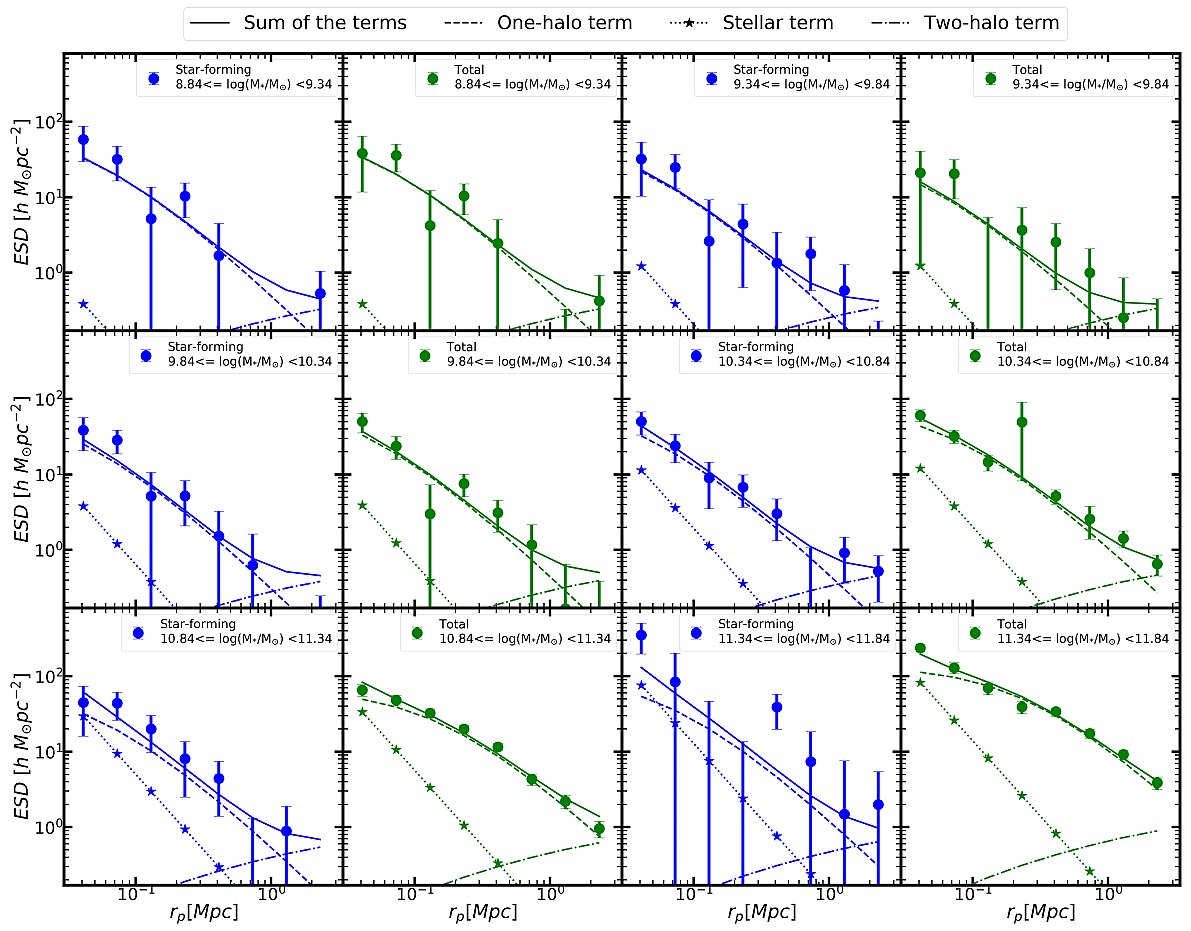}
\centering
\caption{The same as in Figure \ref{fig_lens} but for SDSS shear catalog.}\label{fig_lens2}
\end{figure*}
As an independent test, we also use a different shear catalog \citep{Luo2017ApJ} 
based on SDSS DR7 \citep{Abazajian-09}. The difference between the two samples is twofold. 
First, the shape measurement method here uses a traditional second-moment estimator to 
evaluate the ellipticity; secondly, the coverage of the SDSS DR7 is much larger 
than the DECaLS region overlapping with our lens samples. 
There are 190,730 galaxies with $M_*\ge 10^{8.8}\Msun$ in the star-forming sample, 
and 445,135 galaxies with $M_*\ge 10^{8.8}\Msun$ in the total sample.
However, the deeper imaging, together with the
PDF-symmetrization method, gives much smaller statistical errors than the 
SDSS DR7 catalog. We repeat the modeling as described above to extract the halo mass.
The results are consistent with those obtained from the DECaLS shear data, but   
with larger error bars as shown in Figure \ref{fig_eff}.  The ESDs and best-fitting 
results obtained from the SDSS shear catalog are presented in Figure \ref{fig_lens2}.
\FloatBarrier

\section{The IllustrisTNG Simulation used for comparison}\label{TNG_res}

The IllustrisTNG simulations are run with $\rm AREPO$\citep{Springel2010}, 
which implements a moving-mesh technique. The simulations include subgrid physical 
models for gas cooling, star formation, metal enrichment, and stellar and AGN feedback. 
In this paper, we use the simulation TNG100-1, which has a box size of $\sim 100\Mpc$. 
The dark matter particle mass is $7.5\times10^6\Msun$ and the 
average gas cell mass is about $1.39\times10^6\Msun$. Galaxies with 
$M_*\ge10^{8.8}\Msun$ are well resolved in the simulation. 
Central galaxies are defined as the most massive galaxies
in their host halos.  Galaxy stellar mass is the sum of all 
stellar particles within the gravitationally bound substructure, 
and the SFR is calculated from all gas cells in the same region.
The halo mass, directly taken from the simulation, is
the mass contained in spherical regions, within each of which the mean
mass density is 200 times of the cosmic mean matter density. 

\begin{figure}
\includegraphics[scale=0.355]{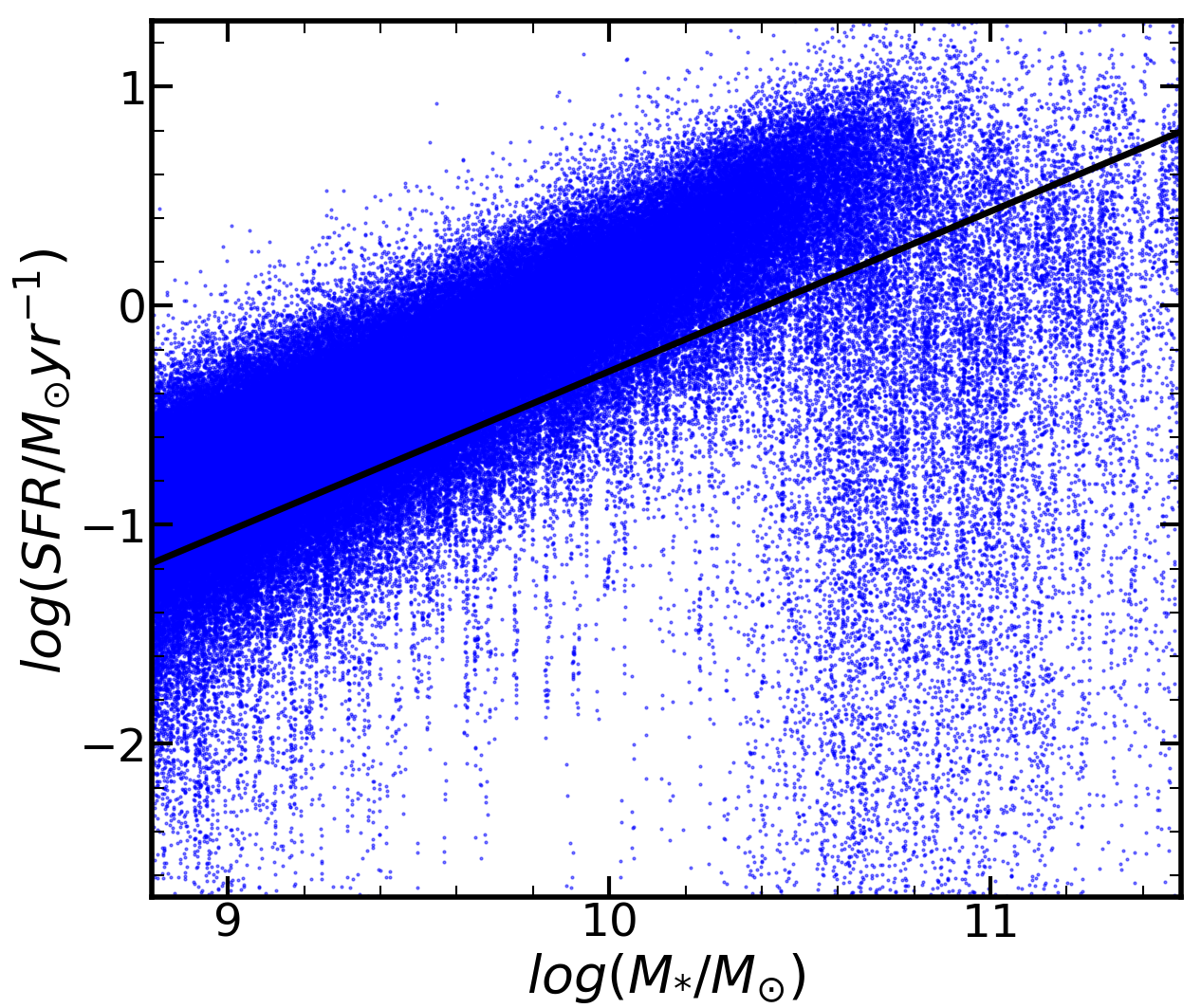}
\centering
\caption{The SFR$-M_*$ diagram for model galaxies. The figure shows 
the result for central galaxies in IllustrisTNG. The solid line in the panel is 
the demarcation line used to identify star-forming galaxies, which is the same 
as the one used in the observation. Note that a large fraction of the 
galaxies have very low SFR. These galaxies fall outside the boundary of the plot.}
\label{fig_sm}
\end{figure}

In Figure \ref{fig_sm}, we show the SFR-stellar mass relation
for central galaxies in the simulation. 
The simulation can well reproduce the star-forming main sequence, 
and most star-forming galaxies in the stellar mass range plotted 
are well above the demarcation line used in our observational data. 
We thus decide to use the same demarcation criteria to identify 
star-forming galaxies. The efficiency as a function of stellar mass for 
both star-forming galaxies and total galaxy populations is presented 
in Figure \ref{fig_eff}. 

\end{appendix}

\end{document}